\documentclass{article}

\usepackage{microtype}
\usepackage{graphicx}
\usepackage{booktabs} 

\usepackage{hyperref}

\usepackage{todonotes}

\usepackage{subcaption}
\usepackage{multirow}
\usepackage{xcolor}

\usepackage{amsmath}
\usepackage{amsthm}
\usepackage{amssymb}

\usepackage{overpic}

\DeclareMathOperator*{\dd}{d}
\DeclareMathOperator*{\id}{id}

\usepackage{cleveref}

\newtheorem{prop}{Proposition}

\usepackage[accepted]{icml2020}

\icmltitlerunning{Forecasting Sequential Data Using Consistent Koopman Autoencoders}

\begin{document}

\twocolumn[
\icmltitle{Forecasting Sequential Data Using Consistent Koopman Autoencoders}



\icmlsetsymbol{equal}{*}

\begin{icmlauthorlist}
\icmlauthor{Omri Azencot}{equal,to}
\icmlauthor{N. Benjamin Erichson}{equal,Berkeley}
\icmlauthor{Vanessa Lin}{Berkeley}
\icmlauthor{Michael W. Mahoney}{Berkeley}
\end{icmlauthorlist}

\icmlaffiliation{to}{Department of Mathematics at UC Los Angeles, CA, USA.}
\icmlaffiliation{Berkeley}{ICSI and Department of Statistics at UC Berkeley, CA, USA.}

\icmlcorrespondingauthor{Omri Azencot}{azencot@math.ucla.edu}
\icmlcorrespondingauthor{N. Benjamin Erichson}{erichson@berkeley.edu}


\vskip 0.3in
]



\printAffiliationsAndNotice{\icmlEqualContribution} 

\begin{abstract}

Recurrent neural networks are widely used on time series data, yet such models often ignore the underlying physical structures in such sequences. A new class of physics-based methods related to Koopman theory has been introduced, offering an alternative for processing nonlinear dynamical systems. In this work, we propose a novel Consistent Koopman Autoencoder model which, unlike the majority of existing work, leverages the forward and backward dynamics. Key to our approach is a new analysis which explores the interplay between consistent dynamics and their associated Koopman operators. Our network is directly related to the derived analysis, and its computational requirements are comparable to other baselines. We evaluate our method on a wide range of high-dimensional and short-term dependent problems, and it achieves accurate estimates for significant prediction horizons, while also being robust to noise.

\end{abstract}

\vspace{-5mm}
\section{Introduction}

Sequential data processing and forecasting is a fundamental problem in the engineering and physics sciences. Recurrent Neural Networks (RNNs) provide a powerful class of models for these tasks, designed to learn long-term dependencies via their hidden state variables. However, training RNNs over long time horizons is notoriously hard \citep{pascanu2013difficulty} due to the problem of exploding and vanishing gradients~\citep{bengio1994learning}. Several approaches have been proposed to mitigate this issue using unitary hidden-to-hidden weight matrices \citep{arjovsky2016unitary} or analyzing stability properties~\citep{miller2018stable}, among other solutions. Still, attaining long-term memory remains a challenge and moreover, modeling short-term dependencies might also be affected due to the limited expressivity of unitary RNNs~\citep{kerg2019non}.

Another shortcoming of traditional RNNs is their difficulty to incorporate high-level constraints into the model. In this context, physics-based methods have been proposed, relating RNNs to dynamical systems \citep{sussillo2013opening} or differential equations \citep{chang2018antisymmetricrnn}. This point of view allows one to construct models which enjoy high-level properties such as time invertibility via Hamiltonian \citep{greydanus2019hamiltonian} or Symplectic \citep{chen2020symplectic,zhong2020symplectic} networks. Other techniques suggested reversible RNNs~\citep{mackay2018reversible,chen2018neural} to alleviate the large memory footprints RNNs induce during training. In this work, we advocate that modeling time-series data which exhibit strong short-term dependencies can benefit from \emph{relaxing} the strict stability as well as time invertibility requirements. 

An interesting physics motivated alternative for analyzing time series data has been introduced in Koopman-based models~\citep{takeishi2017learning,morton2018deep,morton2019deep,li2020learning}. Koopman theory is based on the insight that a nonlinear dynamical system can be fully encoded using an operator that describes how scalar functions propagate in time. The Koopman operator is \emph{linear}, and thus preferable to work with in practice, as tools from linear algebra and spectral theory can be directly applied. While Koopman's theory \citep{koopman1931hamiltonian} was established almost a century ago, significant advances have been recently accomplished in the theory and methodology with applications in the fluid mechanics~\citep{mezic2005spectral} and geometry processing \citep{ovsjanikov2012functional} communities.

The Koopman operator maps between function spaces and thus it is infinite-dimensional and can not be represented on a computer. Nevertheless, most machine learning approaches hypothesize that there exists a data transformation under which an approximate finite-dimensional Koopman operator is available. Typically, this map is represented via an autoencoder network, embedding the input onto a low-dimensional latent space. In that space, the Koopman operator is approximated using a linear layer that encodes the dynamics~\citep{takeishi2017learning}. The main advantage of this framework is that the resulting models are easy to analyze, and they allow for accurate prediction of short-term dependent data. Specifically, predicting forward or backward in time can be attained via subsequent matrix-vector products between the Koopman matrix and the latent observation. Similarly, stability features can be analyzed and constrained via the operator spectrum.

Based on the theory of Koopman, we propose a new model for forecasting high-dimensional time series data. In contrast to previous approaches, we assume that the \emph{backward} map exists. That is, the system from a future to current time can be properly defined. While not all systems exhibit this feature (e.g., diffusive systems), there are many practical cases where this assumption holds. We investigate the interplay between the forward and backward maps and their consistency in the discrete and continuous space settings. Our work can be viewed as relaxing both reversibility and stability requirements, leading to higher expressivity and improved forecasts. Ideas close to ours have been applied to training Generative Adversarial Networks~\citep{zhu2017unpaired,hoffman2018cycada}. However, to the best of our knowledge, our work is first to establish the link between consistency of latent variables and dynamical systems.

\subsection{Background and Problem Setup}
\label{subsec:pbm_def}

In what follows, we focus on dynamical systems that can be described by a time-invariant model
\begin{align} \label{eq:disc_dyn}
	z_{k+1} = \varphi(z_k) \ , \quad z \in \mathcal{M} \subset \mathbb{R}^m \ ,
\end{align}
where $z_k$ denotes the state of the system at time $k \in \mathbb{N}$. The map $\varphi : \mathcal{M} \rightarrow \mathcal{M}$ is a (potentially non-linear) update rule on a finite dimensional manifold $\mathcal{M}$, pushing states from time $k$ to time $k+1$. The above model assumes that future states depend only on the current state $z_k$ and not on information from a sequence of previous states.

\begin{figure}[!t]
	\centering
	\vspace{+0.1cm}
	\begin{overpic}[width=1\linewidth]{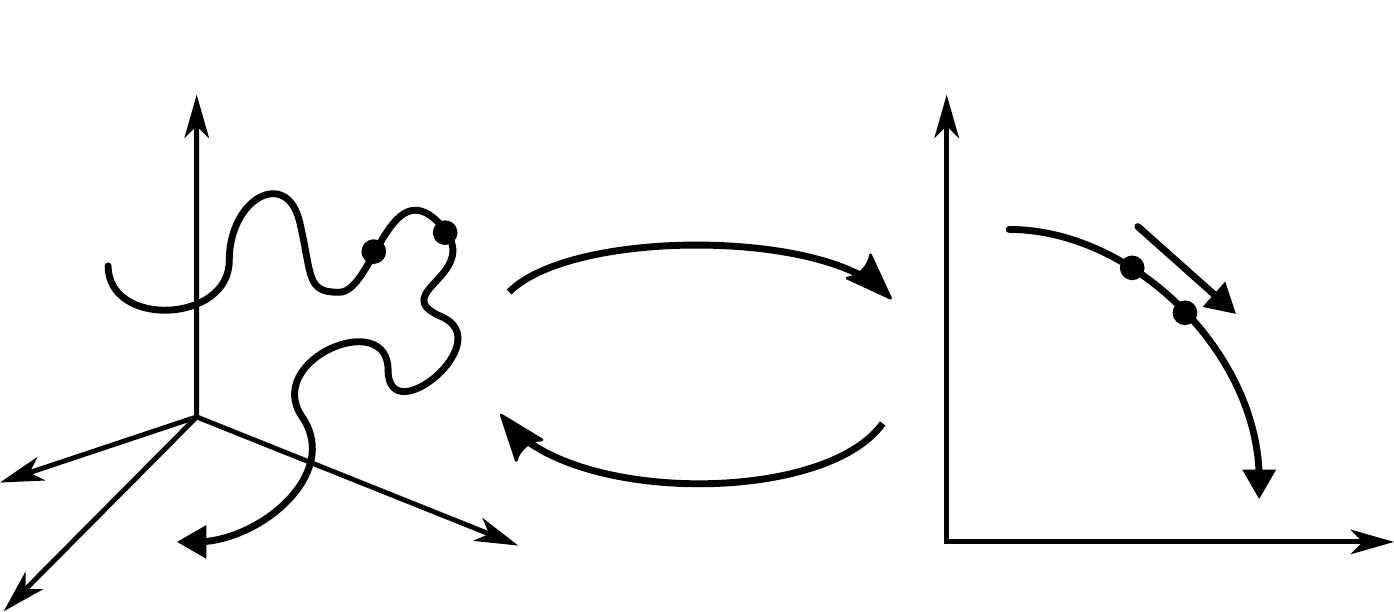}
		\put(45,30){{\small encode}} \put(45,3){{\small decode}}
		\put(10,40){ \small Nonlinear} \put(75,40) { \small Linear}
		\put(25,32){{$\varphi(z_k)$}} \put(85,28){{$\mathcal{K}_\varphi f(z_k)$}}
	\end{overpic}	
	\caption{Illustration of the data transformation that maps the high-dimensional states $z_k$ which evolve on a nonlinear trajectory via $\varphi$ to a new space where the dynamics are linear and given by $\mathcal{K}_\varphi$.}
	\label{fig:coordinate_trans}
\end{figure}
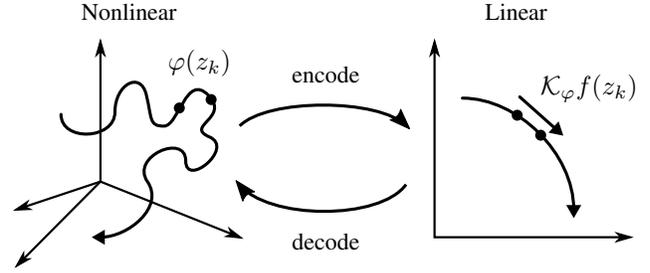

In order to predict future states, one might be tempted to train a neural network which learns an approximation of the map $\varphi$. However, the resulting model ignores prior knowledge related to the problem, and it can be challenging to analyze. Alternatively, our approach is based on a data transformation of the states $z_k$ for which the corresponding latent variables $q_k$ evolve on a linear path, as illustrated in Figure~\ref{fig:coordinate_trans}. Our method allows one to integrate into the training process physics priors such as the relation between the forward and backward maps. Interestingly, Koopman theory suggests that indeed, there exists such a data transformation for any nonlinear dynamical system, so that the evolution of states can be fully represented by a linear map. 

\begin{figure}[!t]
    \centering
    \begin{overpic}[width=\linewidth]{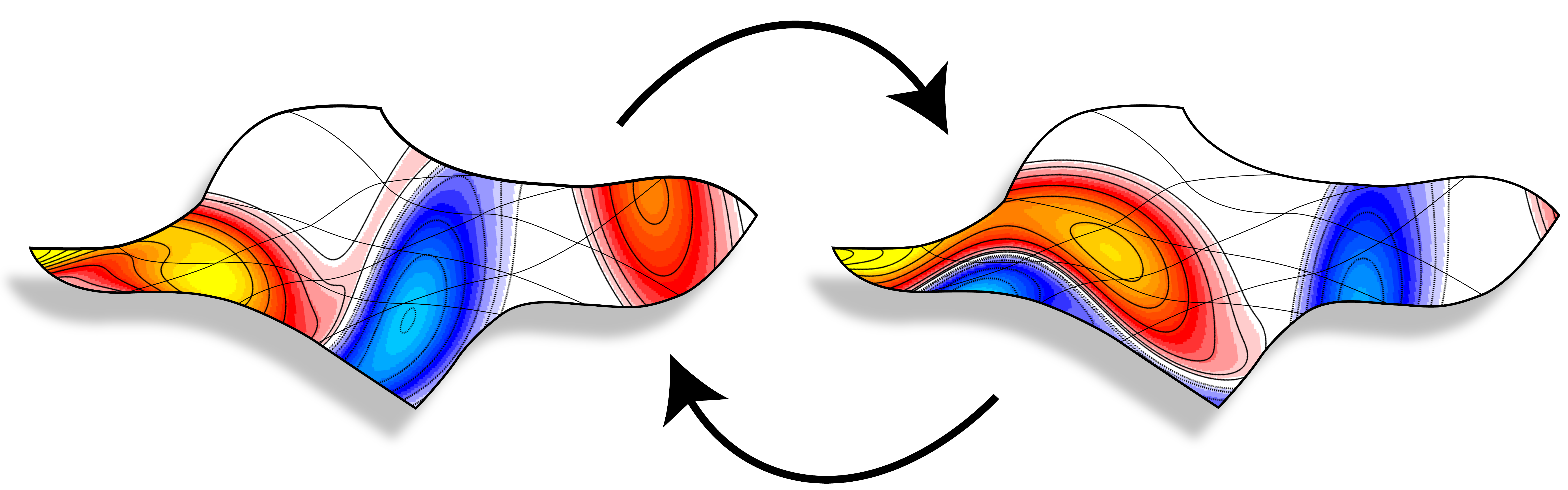}
        \put(32,25) {$\mathcal{K}_\varphi$} \put(65,1) {$\mathcal{U}_\psi$}
    \end{overpic}
    \caption{Our analysis and computational pipeline consider the forward and backward dynamics. Here, the operator $\mathcal{K}_\varphi$ acts on a fluid flow field, detailing the evolution by pulling-back information from a future state. The operator $\mathcal{U}_\psi$ acts in the reverse direction, i.e., prescribing the backward evolution by pushing-forward information from a past observation.}
    \label{fig:fwd_bwd_koopman}
\end{figure}

More formally, the dynamics $\varphi$ induces an operator $\mathcal{K}_\varphi$ that acts on scalar functions $f: \mathcal{M} \rightarrow \mathbb{R} \in \mathcal{F}$, with $\mathcal{F}$ being some function space on $\mathcal{M}$~\citep{koopman1931hamiltonian}. The \emph{Koopman operator} is defined by
\begin{align}
	\mathcal{K}_\varphi f(z) = f \circ \varphi(z) \ , 
\end{align}
i.e., the function $f$ is composed with the map $\varphi$. Essentially, the operator prescribes the evolution of a scalar function by pulling-back its values from a future time. In other words, $\mathcal{K}_\varphi f$ at $z_k$ is the value of $f$ evaluated at the future state $z_{k+1}$. Hence, the Koopman operator is also commonly known as the pull-back operator. Also, it is easy to show that $\mathcal{K}_\varphi$ is \emph{linear} for any $\alpha, \beta \in \mathbb{R}$
\begin{align*}
\mathcal{K}_\varphi (\alpha f + \beta g) &= (\alpha f + \beta g) \circ \varphi \\
&= \alpha f \circ \varphi + \beta g \circ \varphi = \alpha \mathcal{K}_\varphi f + \beta \mathcal{K}_\varphi g \ .
\end{align*}
Finally, we assume that the backward dynamics $\psi$ exists, and we denote by $\mathcal{U}_\psi$ the associated Koopman operator. In Fig.~\ref{fig:fwd_bwd_koopman}, we show an illustration of our setup.

Unfortunately, $\mathcal{K}_\varphi$ is infinite-dimensional. Nevertheless, the key assumption in most of the practical approaches is that there exists a transformation $\chi$ whose conjugation with $\mathcal{K}_\varphi$ leads to a \emph{finite-dimensional} approximation which encodes ``most'' of the dynamics. Formally, 
\begin{align}
	C = \chi \circ \mathcal{K}_\varphi \circ \chi^{-1} \ , \quad C \in \mathbb{R}^{\kappa \times \kappa} \ ,
\end{align}
i.e., $\chi$ and its inverse extract the crucial structures from $\mathcal{K}_\varphi$, yielding an approximate Koopman matrix $C$. Similarly, we denote by $D = \chi \circ \mathcal{U}_\psi \circ \chi^{-1}$ the approximate backward system. The main focus in this work is to find the matrices $C$ and $D$, and a nonlinear transformation $\chi$ such that the underlying dynamical system is recovered well.

We assume that we are given scalar observations of the dynamics $\{ f_k : \mathcal{M} \rightarrow \mathbb{R} \}_{k=1}^n$ such that $f_{k+1} = f_k \circ \varphi +  r_k$,
where the function $r_k \in \mathcal{F}$ represents deviation from the true dynamics due to e.g., measurement errors or missing values. We focus on the case where $\varphi$ and $\psi$ are generally unknown, and our goal is to predict future observations from the given ones. Namely,
\begin{align} \label{eq:multistep_pushfwd}
	f_{k+l} = f_k \circ \varphi^l \ , \quad l = 1,2,... \ ,
\end{align}
where $\varphi^l$ means we repeatedly apply the dynamics. In practice, as $C$ approximates the system, we exploit the relation $\chi^{-1} \circ C^l \circ \chi (f_k) \approx f_k \circ \varphi^l$ to produce further predictions. That is, the matrix $C$ fully determines the forward evolution of the input observation $f_k$.

\subsection{Main Contributions}

Our main contributions are as follows.
\vspace{-3mm}
\begin{itemize}
	\item We developed a Physics Constrained Learning (PCL) framework based on Koopman theory and consistent dynamics for processing complex time series data.
	
	\item Our model is efficient and effective and its features include accurate predictions, time reversibility, and stable behavior even over long time horizons.
	
	\item We evaluate on high-dimensional clean and noisy systems including the pendulum, cylinder flow, vortex flow on a curved domain, and climate data, and we achieve exceptionally good results with our model.
\end{itemize}

\section{Related Work}

Modeling dynamical systems from Koopman's point-of-view has gained increasing popularity in the last few years~\citep{mezic2005spectral}. An approximation of the Koopman operator can be computed via the Dynamic Mode Decomposition (DMD) algorithm~\citep{schmid2010dynamic}. While many extensions of the original algorithm have been proposed, most related to our approach is the work of~\cite{azencot2019consistent} where the authors consider the forward and backward dynamics in a non-neural optimization setting. A network design similar to ours was proposed by \cite{lusch2018deep}, but without our analysis, back prediction and consistency terms. Other techniques minimize the residual sum of squares~\citep{takeishi2017learning,morton2018deep}, promote stability~\citep{erichson2019physics,pan2020physics}, use a variational approach ~\cite{mardt2018vampnets}, or use graph convolutional networks~\citep{li2020learning}.
%

Sequential data are commonly processed using RNNs~\citep{elman1990finding,graves2012supervised}. The main difference between standard neural networks and RNNs is that the latter networks maintain a hidden state which uses the current input and previous inner states. Variants of RNNs such as Long Short Term Memory \citep{hochreiter1997long} and Gated Recurrent Unit \citep{cho2014learning} have achieved groundbreaking results on various tasks including language modeling and machine translation, among others. Still, training RNNs involves many challenges and a recent trend in machine learning focuses on finding new interpretations of RNNs based on dynamical systems theory~\citep{laurent2016recurrent,miller2018stable}. 

Several physics-based models have been recently proposed. Based on Lagrangian mechanics theory, \citet{lutter2018deep} encoded the Euler--Lagrange equations into their network to attain physics plausibility and to alleviate poor generalization of deep models. Other methods attempt to learn conservation laws from data and their associated Hamiltonian representation, leading to exact preservation of energy \citep{greydanus2019hamiltonian} and better handling of stiff problems \citep{chen2020symplectic}. To deal with the limited expressivity of unitary RNNs, \citet{kerg2019non} suggested to employ the Schur decomposition to their connectivity matrices. By considering the normal and nonnormal components, their network allows for transient expansion and compression, leading to improved results on tasks which require continued computations across timescales. 

\begin{figure*}[!t]
	\centering
	\begin{overpic}[width=0.99\linewidth]{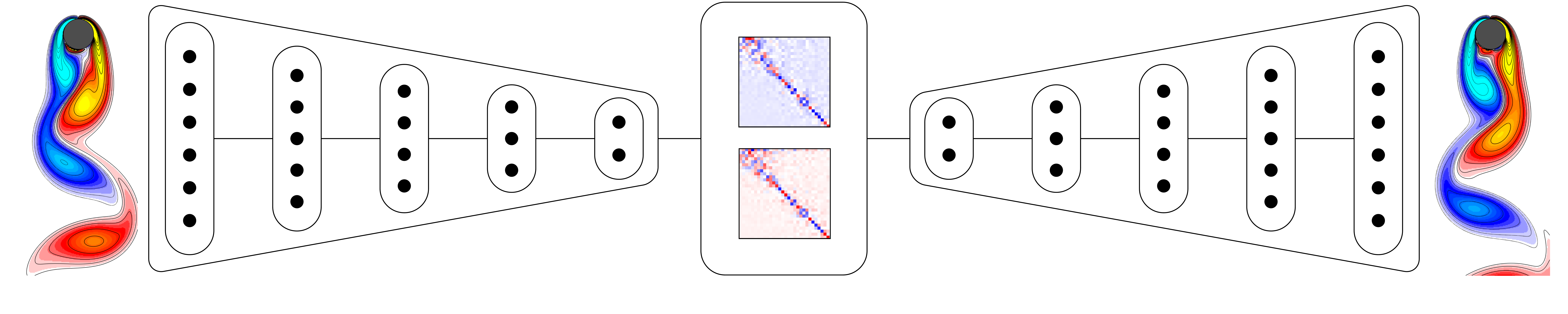}
		\put(45.5,18) {$C$} \put(53,3) {$D$} \put(4,-1) {$f_k$} \put(94,-1) {$f_{k+1}$} \put(25,-1) {$\chi_e | \chi_d$} \put(74,-1) {$\chi_d | \chi_e$}
	\end{overpic}
	\vspace{+0.2cm}
	\caption{Our network takes observations $f_k$, and it learns a latent representation of them via an autoencoder architecture $\chi_e$ and $\chi_d$. Then, the latent variables are propagated forward and backward in time using the linear layers $C$ and $D$, respectively. We emphasize that all the above connections are bi-directional and so information can flow freely from left to right or right to left.}
	\label{fig:net_arch}
\end{figure*}
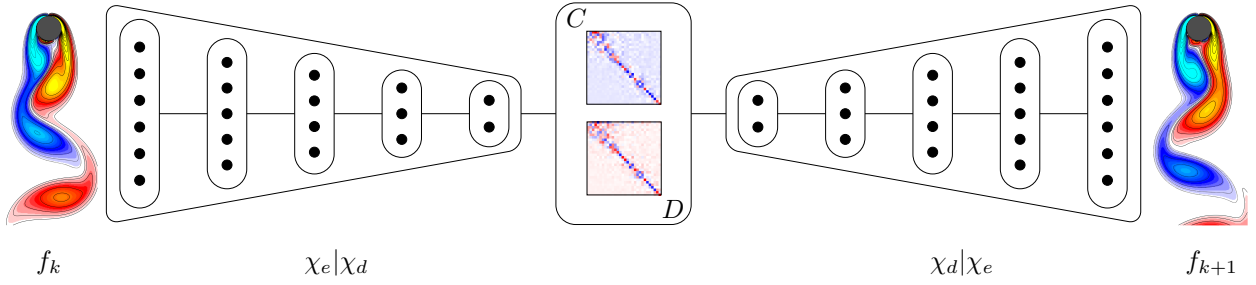

\section{Method}
\label{sec:mtd}

In what follows, we describe our 
PCL 
model which we use to handle time series data. Similar to other methods, we model the transformation $\chi$ above via an encoder $\chi_e$ and a decoder $\chi_d$. Our approach differs from other Koopman-based techniques \cite{lusch2018deep,takeishi2017learning,morton2018deep} in two key components. First, in addition to modeling the forward dynamics, our network also takes into account the \emph{backward} system. Second, we require that the resulting forward and backward Koopman operators are \emph{consistent}. 

\subsection{Autoencoding Observations} Given a set of observations $F = \{ f_k \}_{k=1}^n$ as defined in Sec.~\ref{subsec:pbm_def}, we design an autoencoder (AE) to embed our inputs in a low-dimensional latent space using a nonlinear map $\chi_e$. The decoder $\chi_d$ map allows to reconstruct latent variables in the spatial domain. To train the AE, we define
\begin{align}
    \tilde{f} &= \chi_d \circ \chi_e(f) \ , \\
    \label{eq:loss_id}
    \mathcal{E}_\mathrm{id} &= \frac{1}{2n} \sum_{k=1}^n \| f_k - \tilde{f}_k \|_2^2 \ ,
\end{align}
i.e., $\tilde{f}$ is the reconstructed version of $f$, and $\mathcal{E}_\mathrm{id}$ derives the optimization to obtain an AE such that $\chi_d \circ \chi_e \approx \id$. We note that the specific requirements from $\chi_e$ and $\chi_d$ are problem dependent, and we detail the particular design we use in the appendix.

\subsection{Backward Dynamics}

In general, a dynamical system $\varphi$ prescribes a rule to move forward in time. There are numerous practical scenarios where it makes sense to consider the backward system, i.e., $\psi : z_k \rightarrow z_{k-1}$. For instance, the Euler equation, which describes the motion of an inviscid fluid, is invariant to sign changes in its time parameter (see Sec.~\ref{sec:experiments} for an example). Previous approaches incorporated the backward dynamics into their model as in bi-directional RNN~\cite{schuster1997bidirectional}. However, the inherent nonlinearities of a typical neural network make it difficult to constrain the forward and backward models. To this end, a few approaches were recently proposed~\cite{greydanus2019hamiltonian,chen2020symplectic} where the obtained dynamics are reversible by construction due to the leapfrog integration. In comparison, most existing Koopman-based techniques do not consider the backward system in their modeling or training.

To account for the forward as well as backward dynamics, we incorporate two linear layers with no biases into our network to represent the approximate Koopman operators. As we assume that $\chi_e$ transforms our data into a latent space where the dynamics are linear, we can directly evolve the dynamics in that space. We introduce the following notation
\begin{align}
\label{eq:approx_fwd_pred}
\hat{f}_{k+1} &= \chi_d \circ C \circ \chi_e (f_k) \ , \\
\label{eq:approx_bwd_pred}
\check{f}_{k-1} &= \chi_d \circ D \circ \chi_e (f_k) \ ,
\end{align}
for every admissible $k$. Namely, the Koopman operators $C, D \in \mathbb{R}^{\kappa\times\kappa}$ allow to obtain forward estimates $\hat{f}_{k+1}$, and backward forecasts $\check{f}_{k-1}$.

In practice, we noticed that our models predict as well as generalize better if we employ a \emph{multistep} forecasting, instead of computing one step forward and backward in time. Given a choice of $\lambda_s \in \mathbb{N}$, the total number of prediction steps, we define the following loss terms
\begin{align}
    \label{eq:loss_pred}
\mathcal{E}_{\mathrm{fwd}} &= \frac{1}{2 \lambda_s n} \sum_{l=1}^{\lambda_s} \sum_{k=1}^n \| f_{k+l} - \hat{f}_{k+l} \|_2^2 \ , \\
    \label{eq:loss_bwd_pred}
    \mathcal{E}_{\mathrm{bwd}} &= \frac{1}{2 \lambda_s n} \sum_{l=1}^{\lambda_s} \sum_{k=1}^n \| f_{k-l} - \check{f}_{k-l} \|_2^2 \ ,
\end{align}
where we assume that $f_{k+l}$ and $f_{k-l}$ are provided during training for any $l \leq \lambda_s$, see Eq.~\eqref{eq:multistep_pushfwd}. Also, $\hat{f}_{k+1}$ and $\check{f}_{k-l}$ are obtained by taking powers $l$ of $C$, respectively $D$, in Eqs.~\eqref{eq:approx_fwd_pred} and \eqref{eq:approx_bwd_pred}. We show in Fig.~\ref{fig:net_arch} a schematic illustration of our network design including the encoder and decoder components as well as the Koopman matrices $C$ and $D$. Notice that all the connections are bi-directional, that is, data can flow from left to right \emph{and} right to left.

\textbf{Backward prediction.} Koopman operators and their approximating matrices are linear objects that allow for greater flexibility when compared to other models for time series processing. One consequence of this linearity is that while existing Koopman-based nets \cite{lusch2018deep,takeishi2017learning,morton2018deep} are geared towards forward prediction, their evolution matrix $C$ can be exploited for backward prediction as well. This computation is obtained via the inverse of $C$, i.e.,
\begin{align} \label{eq:approx_fwd_bwd_pred}
    \bar{f}_{k-1} = \chi_d \circ C^{-1} \circ \chi_e(f_k) \ .
\end{align}

However, models that were trained for forward prediction typically produce poor backward predictions as we show in the appendix. In contrast, we note that our model allows for the direct back prediction using the $D$ operator and Eq.~\eqref{eq:approx_bwd_pred}. Thus, while other techniques can technically produce backward predictions, our model supports it by construction.

\subsection{Consistent Dynamics}

The backward prediction penalty $\mathcal{E}_\mathrm{bwd}$ in itself only affects $D$, and it is completely independent of $C$. That is, $C$ will not change due to backpropagating the error in Eq.~\eqref{eq:loss_bwd_pred}. To link between the forward and backward evolution matrices, we need to introduce an additional penalty that promotes consistent dynamics. Formally, we say that the maps $\varphi$ and $\psi$ are \emph{consistent} if $\psi \circ \varphi(f) = f$ for any $f \in \mathcal{F}$. In the Koopman setting, we will show below that this condition is related to requiring that $D C = I_\kappa$, where $I_\kappa$ is the identity matrix of size $\kappa$. However, our analysis shows that in fact the continuous space and discrete space settings differ, yielding related but different penalties. In this work, we will incorporate the following loss to promote consistency
\begin{align} \label{eq:loss_con}
\mathcal{E}_\mathrm{con} &= \sum_{k=1}^\kappa \frac{1}{2k} \| D_{k*} C_{*k} - I_k \|_F^2  + \frac{1}{2k}\| C_{k*} D_{*k} - I_k \|_F^2 \ ,
\end{align}
where $D_{k*}$ and $C_{*k}$ are the upper $k$ rows of $D$ and leftmost $k$ columns of the matrix $C$, and $\| \cdot \|_F$ is the Frobenius norm.

\textbf{Stability.} Recently, stability has emerged as an important component for analyzing neural nets~\cite{miller2018stable}. Intuitively, a dynamical system is stable if nearby points stay close under the dynamics. Mathematically, the eigenvalues of a linear system fully determine its behavior, providing a powerful tool for stability analysis. Indeed, the challenging problem of vanishing and exploding gradients can be elegantly explained by bounding the modulus of the weight matrices' eigenvalues \cite{arjovsky2016unitary}. To overcome these challenges, one can design networks that are stable by construction; see, e.g., \citet{chang2018antisymmetricrnn}, among others.

\begin{figure}[!t]
	\centering
	\begin{subfigure}[t]{0.22\textwidth}
		\centering
		\begin{overpic}[width=1\textwidth]{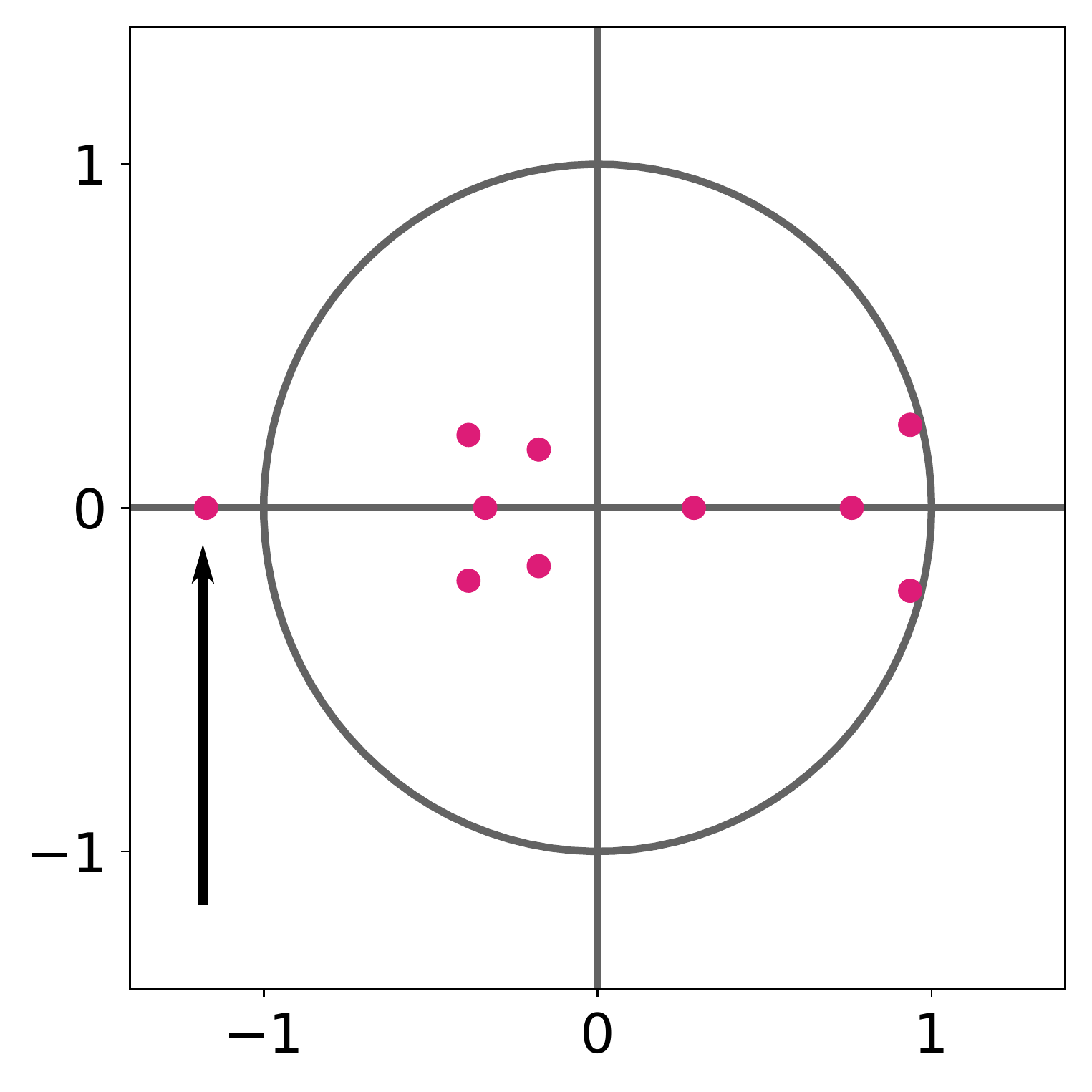} 
			\put(-5,40){\rotatebox{90}{\scriptsize Imaginary}}
			\put(50,-4){\color{black}{\scriptsize Real}}  	
			\put(22,15){\color{black}{\scriptsize $|\lambda| > 1$}}  	
		\end{overpic}\vspace{+0.2cm}		
		
		\caption{Unstable model}
	\end{subfigure}\hspace{-0.25cm}
	~
	\begin{subfigure}[t]{0.22\textwidth}
		\centering
		\begin{overpic}[width=1\textwidth]{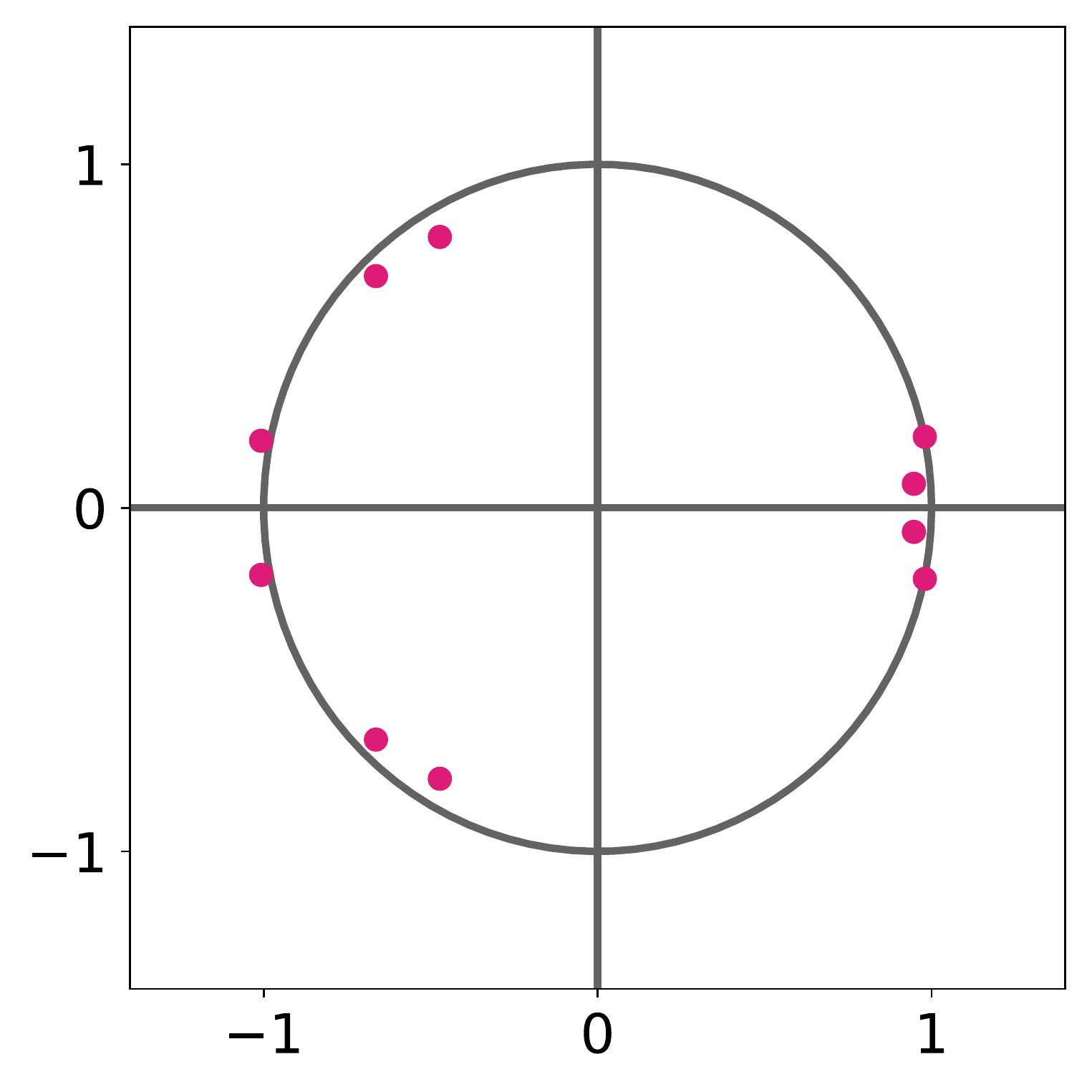} 
			\put(50,-4){\color{black}{\scriptsize Real}}  		
		\end{overpic}\vspace{+0.2cm}			
		\caption{Ours}
	\end{subfigure}
	
	\caption{Koopman operators are linear and thus their spectrum can be investigated. We visualize the eigenvalues of an unregularized model (a) and ours (b) in the complex plane. Indeed, our 
	PCL 
	framework promotes stability, whereas the other model exhibits an eigenvalue with modulus greater than one.}
	\label{fig:fluid_flow1_eigs}
\end{figure}

We, on the other hand, relax the stability constraint and allow for quasi-stable models. In practice, our loss term~\eqref{eq:loss_con} regularizes the nonconvex minimization by promoting the eigenvalues to get closer to the unit circle. The comparison in Fig.~\ref{fig:fluid_flow1_eigs} highlights the stability features our model attains, whereas a non regularized network obtains unstable modes. From an empirical viewpoint, unstable behavior leads to rapidly diverging forecasts, as we show in Sec.~\ref{sec:experiments}. 

\subsection{A Consistent Dynamic Koopman Autoencoder}

Combining all the pieces together, we obtain our 
PCL 
model for processing time series data. Our model is trained by minimizing a loss function whose minimizers guarantee that we achieve a good AE, that predictions through time are accurate, and that the forward and backward dynamics are consistent. We define our loss 
\begin{align} \label{eq:ours_loss}
\mathcal{E} = \lambda_{\mathrm{id}} \mathcal{E}_{\mathrm{id}} + \lambda_{\mathrm{fwd}} \mathcal{E}_{\mathrm{fwd}} + \lambda_{\mathrm{bwd}} \mathcal{E}_{\mathrm{bwd}} + \lambda_{\mathrm{con}} \mathcal{E}_{\mathrm{con}} \ ,
\end{align}
where $\lambda_\mathrm{id}, \lambda_\mathrm{fwd}, \lambda_\mathrm{bwd}, \lambda_\mathrm{con} \in \mathbb{R}^+$ are user-defined positive parameters that balance between reconstruction, prediction and consistency. Finally, $\mathcal{E}_\mathrm{id}, \mathcal{E}_\mathrm{fwd}, \mathcal{E}_\mathrm{bwd}, \mathcal{E}_\mathrm{con}$ are defined in Eqs. \eqref{eq:loss_id}, \eqref{eq:loss_pred}, \eqref{eq:loss_bwd_pred}, and \eqref{eq:loss_con}, respectively.

\section{Consistent Dynamics via Koopman}
\label{sec:inv_koopman}

We now turn to prove a necessary and sufficient condition for a dynamical system to be consistent from a Koopman viewpoint in the continuous space setting. We then show a similar result in the spatial discrete case, yielding a more elaborate condition which we use in practice.

We recall that Koopman operators take inputs and return outputs from a function space $\mathcal{F}$. Thus, many properties of the underlying dynamics can be related to the action of $\mathcal{K}_\varphi$ on every function in $\mathcal{F}$. A natural approach in this case is to consider a \emph{spectral} representation of the associated objects. Specifically, we choose an orthogonal basis for $\mathcal{F}$ which we denote by $\{ \xi_k \}_{k=0}^\infty$ where for any $i, j$ we have
\begin{align}
\langle \xi_i, \xi_j \rangle_\mathcal{M} = \int_\mathcal{M} \xi_i(z) \, \xi_j(z) \dd z  = \delta_{ij} \ ,
\end{align}
with $\delta_{ij}$ being the Kronecker delta function. Under this choice of basis, any function $f \in \mathcal{F}$ can be represented by $f = \sum_k \langle f, \xi_k \rangle_\mathcal{M} \, \xi_k$. Moreover, due to the linearity of $\mathcal{K}_\varphi$ we also have that $\mathcal{K}_{ij} = \langle \xi_i, \mathcal{K}_\varphi \, \xi_j \rangle_{\mathcal{M}}$.

In the following proposition we characterize consistency (invertibility) in the continuous-space case, which is a known result in Ergodic theory~\cite{eisner2015operator}.

\begin{prop} \label{prop:inv_cont_koopman}
	Given a manifold $\mathcal{M}$, the dynamical system $\varphi$ is invertible if and only if for every $i$ and $j$ the Koopman operators $\mathcal{K}_\varphi$ and $\mathcal{U}_\psi$ satisfy $\langle \xi_i, \mathcal{U}_\psi \, \mathcal{K}_\varphi \, \xi_j \rangle_\mathcal{M} = \delta_{ij}$. 
\end{prop}

\paragraph{Proof.} If $\varphi$ is invertible, then the composition $\psi \circ \varphi = \id$ for every $z \in \mathcal{M}$. Thus, 
\begin{align*}
\langle \xi_i, \mathcal{U} \, \mathcal{K} \, \xi_j \rangle_\mathcal{M} &= \int_\mathcal{M} \xi_i(z) \, \xi_j (\psi \circ \varphi (z)) \dd z = \langle \xi_i, \xi_j \rangle_\mathcal{M}\, .
\end{align*}
Conversely, we assume that $\langle \xi_i, \mathcal{U} \, \mathcal{K} \, \xi_j \rangle_\mathcal{M} = \delta_{ij}$ for all $i, j$. It follows that for every $k$ we have $\mathcal{U} \, \mathcal{K} \, \xi_k = \xi_k$ since $\xi_k$ is orthogonal to every $\xi_l, l \neq k$. Let $f$ be some scalar function, then $f(\psi \circ \varphi (z)) = \mathcal{U} \, \mathcal{K} \, f(z) = f(z)$ and thus $ \psi \circ \varphi = \id$.

The main advantage of Prop.~\ref{prop:inv_cont_koopman} is that it can be used directly in a computational pipeline. In particular, if we denote by $C$ and $D$ the $\kappa \times \kappa$ matrices that approximate $\mathcal{K}$ and $\mathcal{U}$, respectively, then the above condition takes the form 
\begin{align} \label{eq:loss_pred_basic}
\frac{1}{2} \| D \, C - I_\kappa \|_F^2 = 0 \ ,
\end{align}
where $I_\kappa$ is and identity matrix of size $\kappa$. This loss term was recently used in~\cite{azencot2019consistent} to construct a robust scheme for computing Dynamic Mode Decomposition operators. However, we prove below that in the \emph{discrete-space} setting, a more elaborate condition is required. In addition, we refer to the work~\cite{melzi2019zoomout} which deals with a closely related setup where isometric maps are being sought. 

\begin{figure}[!b]
	\centering
	\begin{overpic}[width=\linewidth]{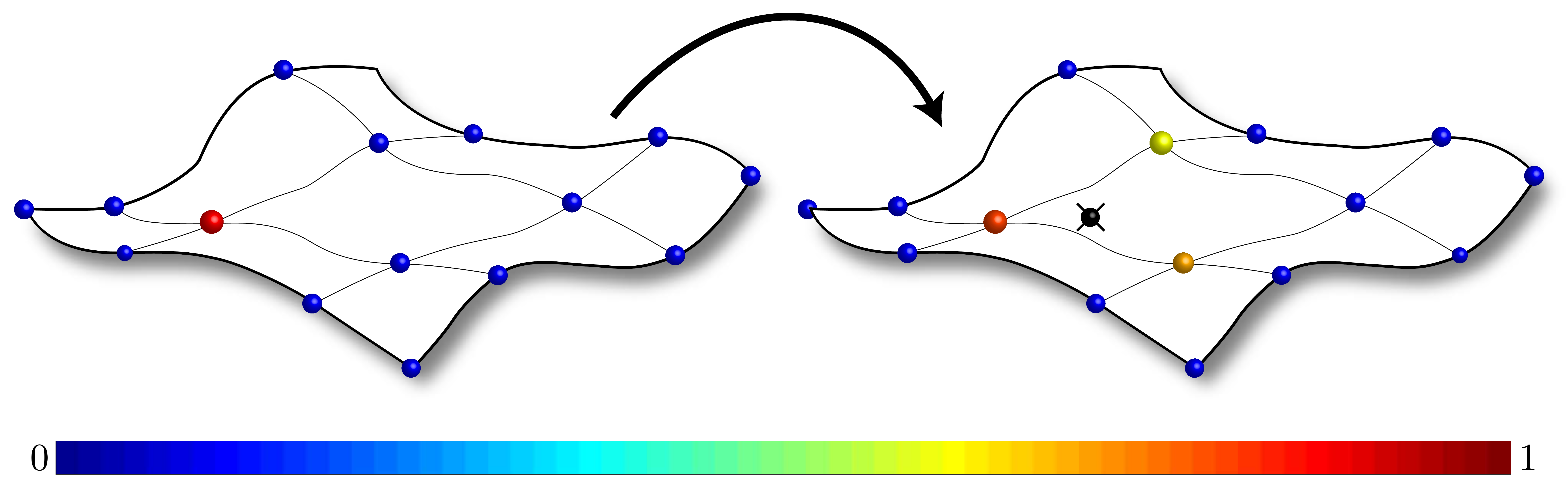}
		\put(35,27) {$P_\varphi$} \put(14,19) {$z$} \put(71,17) {$\varphi(z)$} \put(15,7) {$\delta_z$} \put(65,7) {$h_{\varphi(z)}$}
	\end{overpic}
	\caption{The Kronecker delta $\delta_z$ centered at $z$ is mapped via $P_\varphi$ to the function $h_{\varphi(z)}$ which holds the coefficients of the vertices of $M$ whose combination generates $\varphi(z)$.}
	\label{fig:discrete_manifold}
\end{figure}
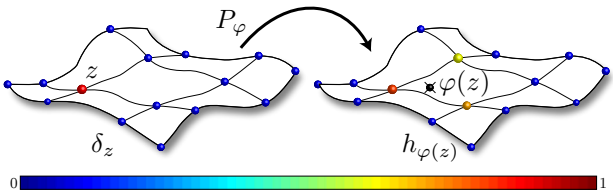

To discretize the above objects, we assume that our manifold $\mathcal{M}$ is represented by the domain $M \subset \mathbb{R}^d$ which is sampled using $m$ vertices. In this setup, scalar functions $f: M \rightarrow \mathbb{R}$ are vectors $f \in \mathbb{R}^m$ storing values on vertices with in-between values obtained via interpolation. A map $\varphi: \mathcal{M} \rightarrow \mathcal{M}$ can be encoded using a matrix $P_\varphi \in \mathbb{R}^{m \times m}$ defined by
\begin{align}
P_\varphi \delta_z = h_{\varphi(z)} \ ,
\end{align}
where $h_x$ is a function that stores the vertices' coefficients such that $h_x^T X = x^T$, with $X \in \mathbb{R}^{m \times d}$ being the spatial coordinates of $M$. Similarly, we denote by $Q_\psi$ the matrix associated with $\psi$, i.e., $ Q_\psi \delta_z = h_{\psi(z)}$. Note that $P$ and $Q$ are in fact discrete Koopman operators represented in the canonical basis. We show in Fig.~\ref{fig:discrete_manifold} an illustration of our spatial discrete setup including some of the notations.

Similar to the continuous setting, we can choose a basis for the function space on $M$. We denote $B \in \mathbb{R}^{m \times m}$ as the matrix that contains the orthogonal basis elements in its columns, i.e., $\langle b_i, b_j \rangle_M = b_i^T b_j = \delta_{ij}$ for every $i, j$. We use this basis to define the matrices $C$ and $D$ by
\begin{align} \label{eq:func_maps}
C = B^T P_\varphi \, B \ , \quad D = B^T Q_\psi \, B \ .
\end{align}
Finally, instead of invertible maps, we consider \emph{consistent} maps. That is, a discrete map $\varphi$ is consistent if for every $z$, we have that $\psi \circ \varphi (z) = z$. Using the above constructions and notations, we are ready to state our main result.

\begin{prop} \label{prop:inv_disc_koopman}
	Given a domain $M$, the map $\varphi$ is consistent if and only if for every $i$ and $j$ the matrices $C$ and $D$ satisfy $\sum_k \frac{1}{2 k} \| D_{k*} \, C_{*k} - I_k \|_F^2 = 0$, where $D_{k*}$ and $C_{*k}$ are the upper $k$ rows of $D$ and leftmost $k$ columns of the matrix $C$.
\end{prop}

\paragraph{Proof.} If $\varphi$ is a consistent map, then $Q \, P \delta_z = \delta_z$ for every $z$ and thus $Q \, P = I$. In addition, for every $k$ we have that 
\begin{align*}
D_{k*} C_{*k} = B_k^T Q  B  B^T P B_k = B_k^T Q  P B_k = B_k^T B_k = I_k,
\end{align*}
where $B_k$ are the first $k$ basis elements of $B$ and $B B^T = I$. Conversely, we assume that $C, D$ are related to some maps $\varphi$ and $\psi$ and are constructed via Eq~\eqref{eq:func_maps}. In addition, the condition $\sum_k \frac{1}{2 k} \| D_{k*} \, C_{*k} - I_k \|_F^2 = 0$ holds. Then $B_k^T Q \, P B_k = I_k$ for every $k$. By induction on $k$, it follows that $Q \, P \, b_k = b_k$ for every $k$, where $b_k$ is the $k$th column of $B$, and thus $Q \, P = I$, as $B$ spans the space of scalar functions on $M$.

\paragraph{Discussion.} Importantly, the indices of the sums in the penalty terms of Eq.~\eqref{eq:loss_con} take values from $[1,\kappa]$, whereas the indices in Prop.~\ref{prop:inv_disc_koopman} reach up to $m$ (the sampling space of $M$). Thus, Eq.~\eqref{eq:loss_con} may be viewed as an approximation of the full term. Nevertheless, this penalty term provides a closer approximation to the full term in comparison to Eq.~\eqref{eq:loss_pred_basic}, for any $\kappa \leq m$. Finally, we emphasize that our discussion also holds for the case $\kappa \ll m$.

\section{Experiments}
\label{sec:experiments}

To evaluate our proposed consistent dynamic Koopman AE, we perform a comprehensive study using various datasets and compare to state-of-the-art Koopman-based approaches as well as other baseline sequential models. Our network minimizes Eq.~\eqref{eq:ours_loss} with a decaying learning rate initially set to $0.01$. We fix the loss weights to $\lambda_\mathrm{id} = \lambda_\mathrm{fwd} = 1$, $\lambda_\mathrm{bwd} = 0.1$, and $\lambda_\mathrm{con} = 0.01$, for the AE, forward forecast, backward prediction and consistency, respectively. We use $\lambda_s = 8$ prediction steps forward and backward in time. We provide additional details in the appendix.
Our code is available at \href{https://github.com/erichson/koopmanAE}{github.com/erichson/koopmanAE}.

\subsection{Baselines}

Our comparison is mainly performed against the state-of-the-art method of \citet{lusch2018deep}, henceforth referred to as the Dynamic AE (DAE) model. Their approach may be viewed as a special case of our network by setting $\lambda_\mathrm{bwd} = \lambda_\mathrm{con} = 0$. While we use this work as a baseline, other models such as~\cite{takeishi2017learning,morton2018deep} could be also considered. The main difference between DAE and the latter techniques is the least squares solution for the evolution matrix $C$ per training iteration. In our experience, this change leads to delicate training procedures, and thus it is less favorable. Unless said otherwise, both models are trained using the same parameters, where DAE does not have the regularizing penalties. 

We additionally compare against a feed-forward model and a recurrent neural network. The feed-forward network simply learns a nonlinear function $\zeta: f_k \rightarrow f_{k+1}$, where during inference we take $\hat{f}_{k+1}$ as input for predicting $\hat{f}_{k+2}$, and so on. The recurrent neural network is similar to the feed-forward model but adds an hidden state $h_k$ such that $h_k = \sigma(U f_k + W h_{k-1} + b)$, and the prediction is obtained via $\hat{f}_k = V h_k + c$. We performed a parameter search when comparing with these baselines.

\begin{figure}[!b]
	\centering
	\begin{overpic}[width=0.47\linewidth]{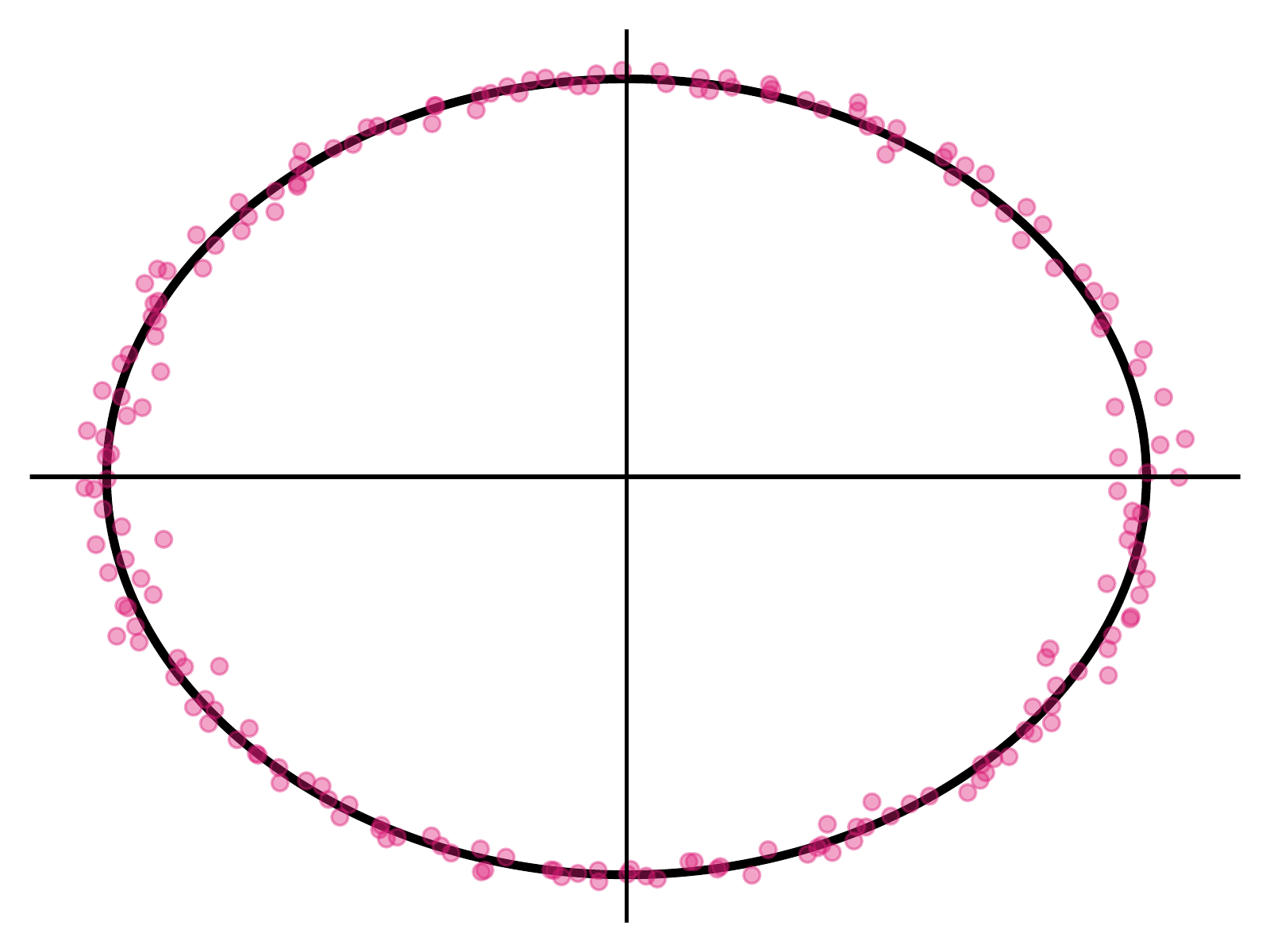}
		\put(41,72){\small $\theta$} \put(95,25){\small $\dot{\theta}$}
	\end{overpic}
	~
	\begin{overpic}[width=0.47\linewidth]{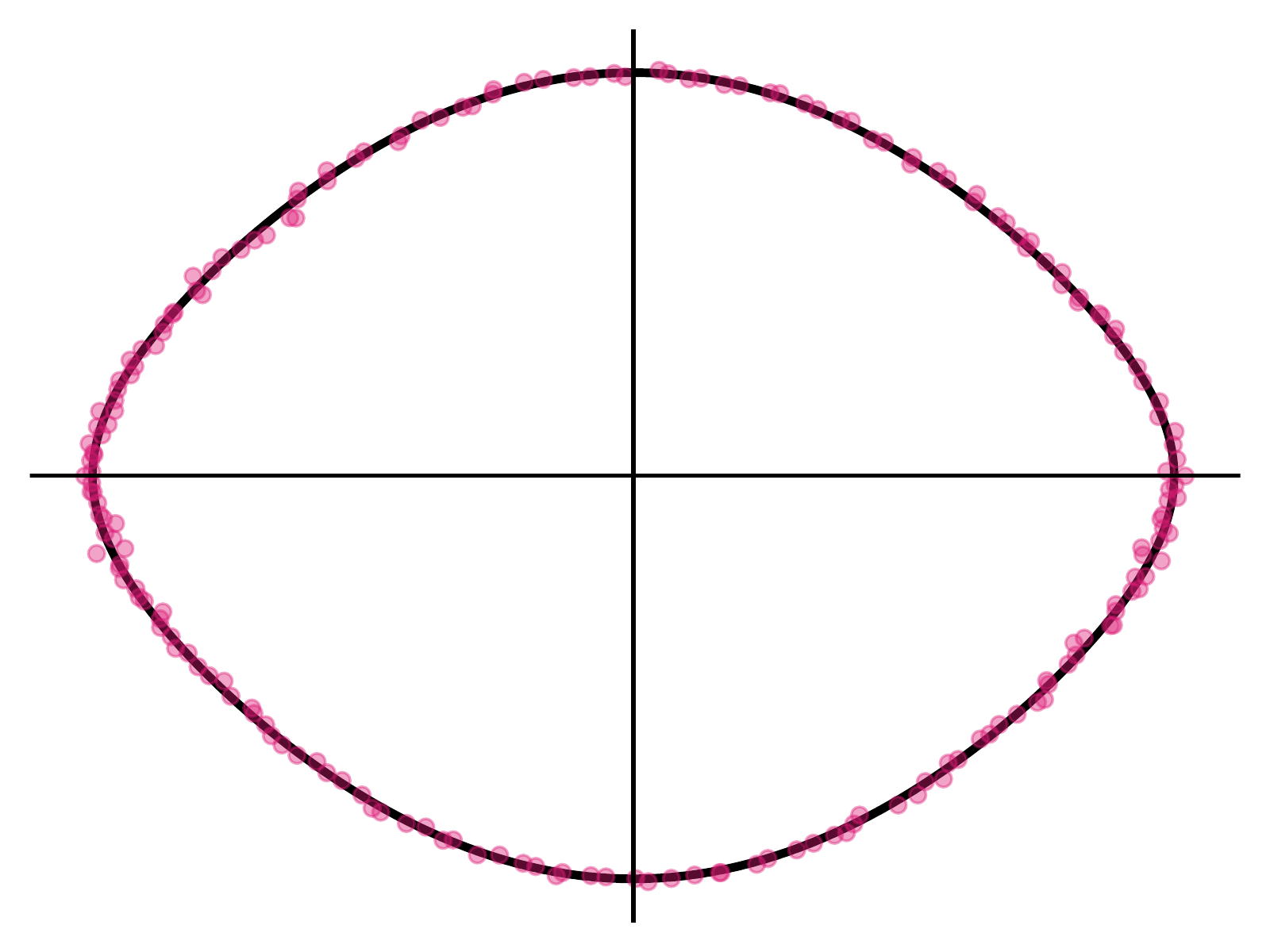}
		\put(41,72){\small $\theta$} \put(95,25){\small $\dot{\theta}$}
	\end{overpic}
	\caption{We show the pendulum's trajectories for the initial conditions of amplitude oscillations $\theta_0=0.8$ (left) and $\theta_0=2.4$ (right). The discrete sampled data points are slightly perturbed.}
	\label{fig:ho_data}
\end{figure}

\begin{figure*}[!t]
	\centering
	\begin{subfigure}[t]{0.45\textwidth}
		\centering
		\begin{overpic}[width=1\textwidth]{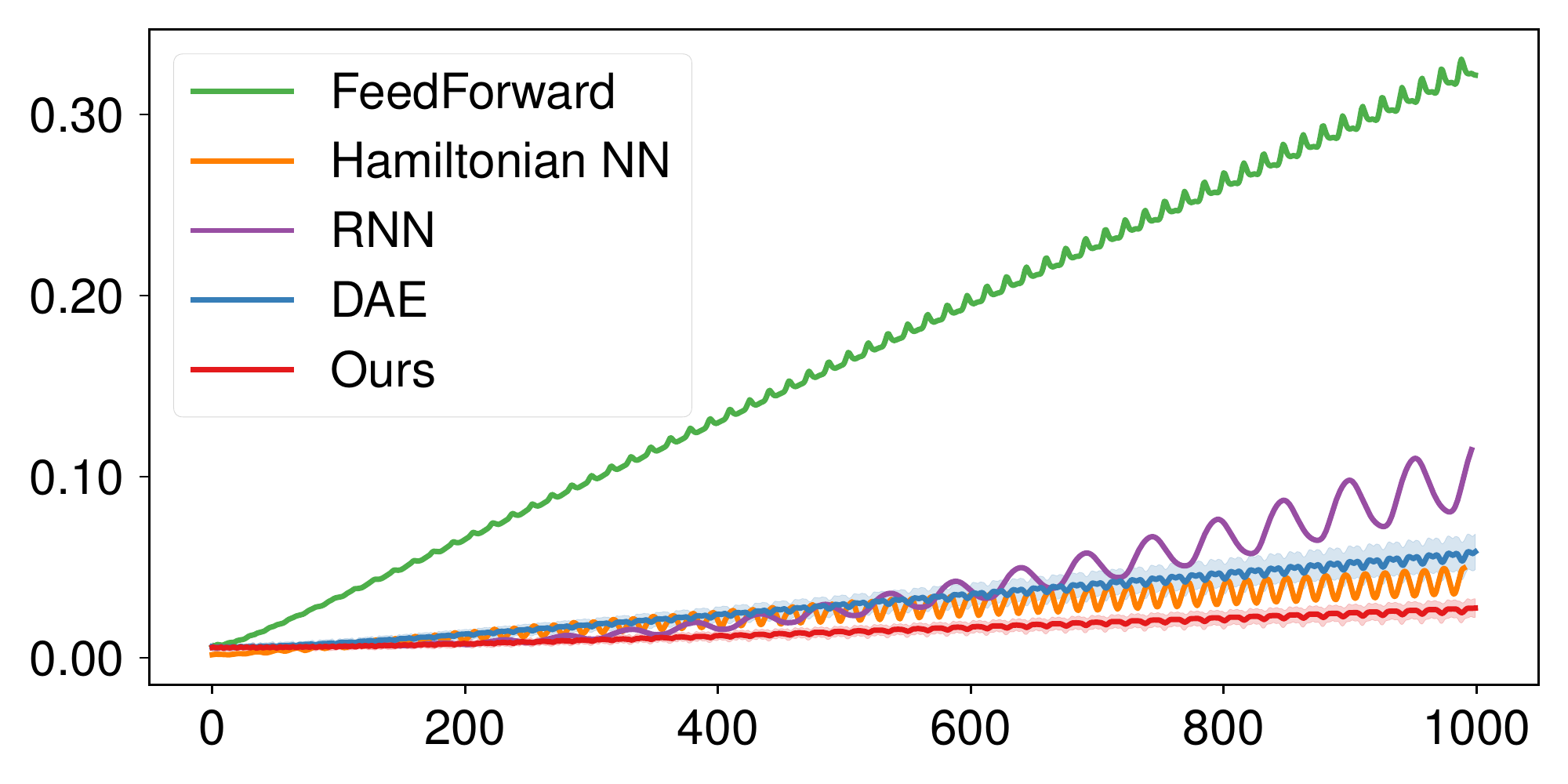} 
			\put(-5,14){\rotatebox{90}{\footnotesize Prediction error}}
		\end{overpic}		
		
		\begin{overpic}[width=1\textwidth]{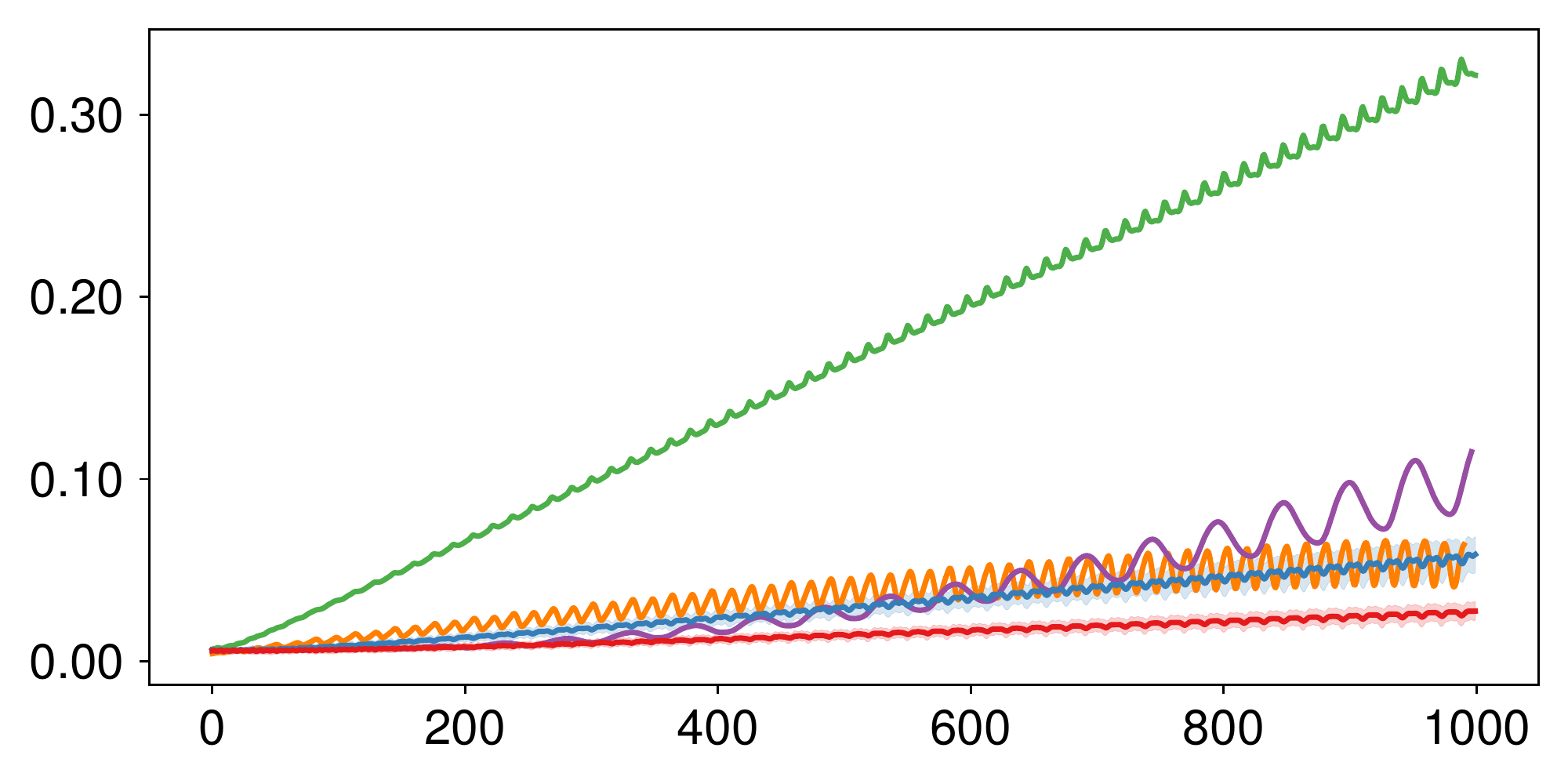} 
			\put(-5,14){\rotatebox{90}{\footnotesize Prediction error}}
			\put(54,-4){\color{black}{\footnotesize $t$}}  		
		\end{overpic}\vspace{+0.32cm}					
		
		\caption{Clean observations}\label{fig:pendulum_1_noise_a}
	\end{subfigure}
	~
	\begin{subfigure}[t]{0.45\textwidth}
		\centering
		\includegraphics[width=1\textwidth]{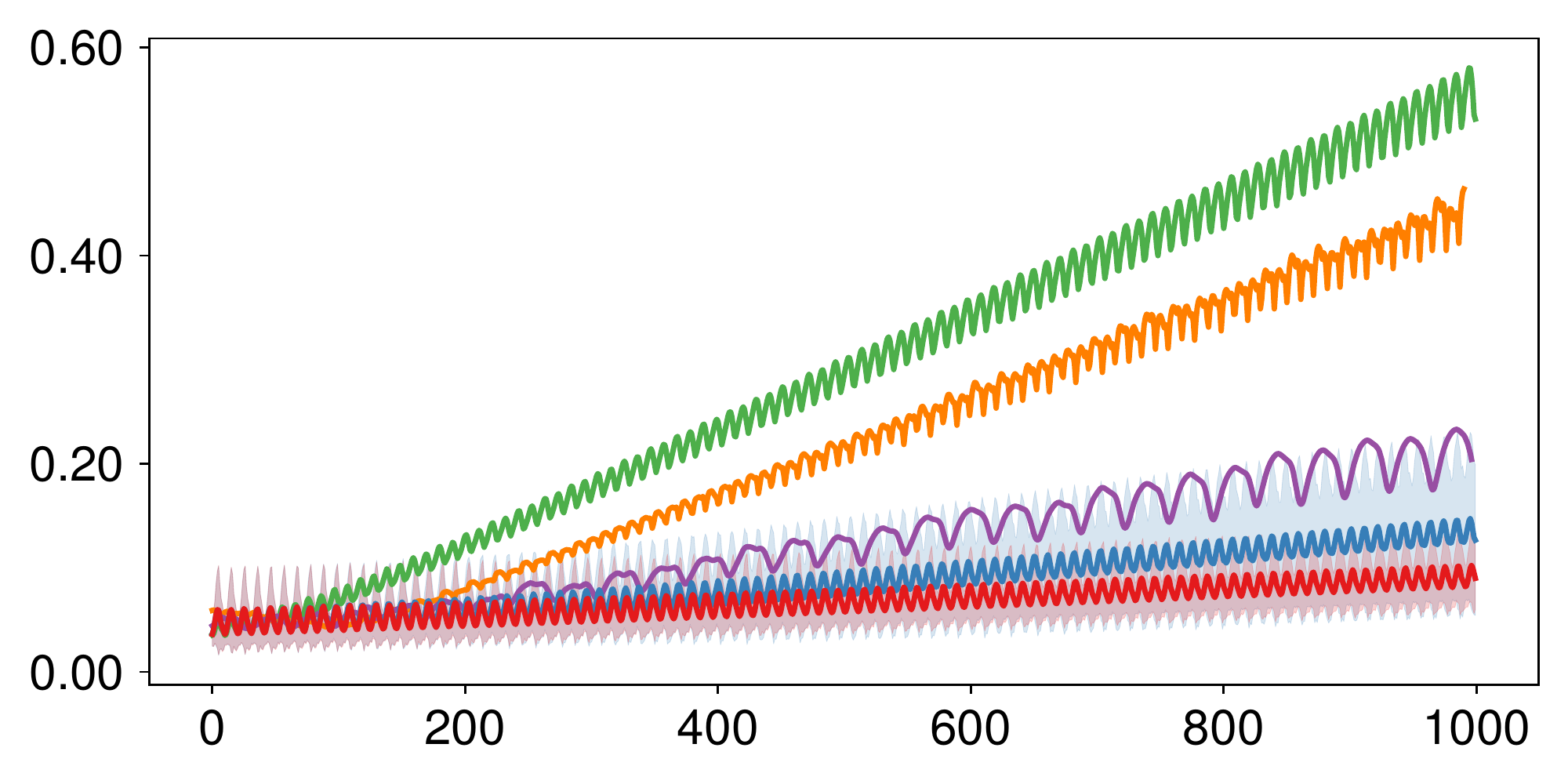}		
		
		\begin{overpic}[width=1\textwidth]{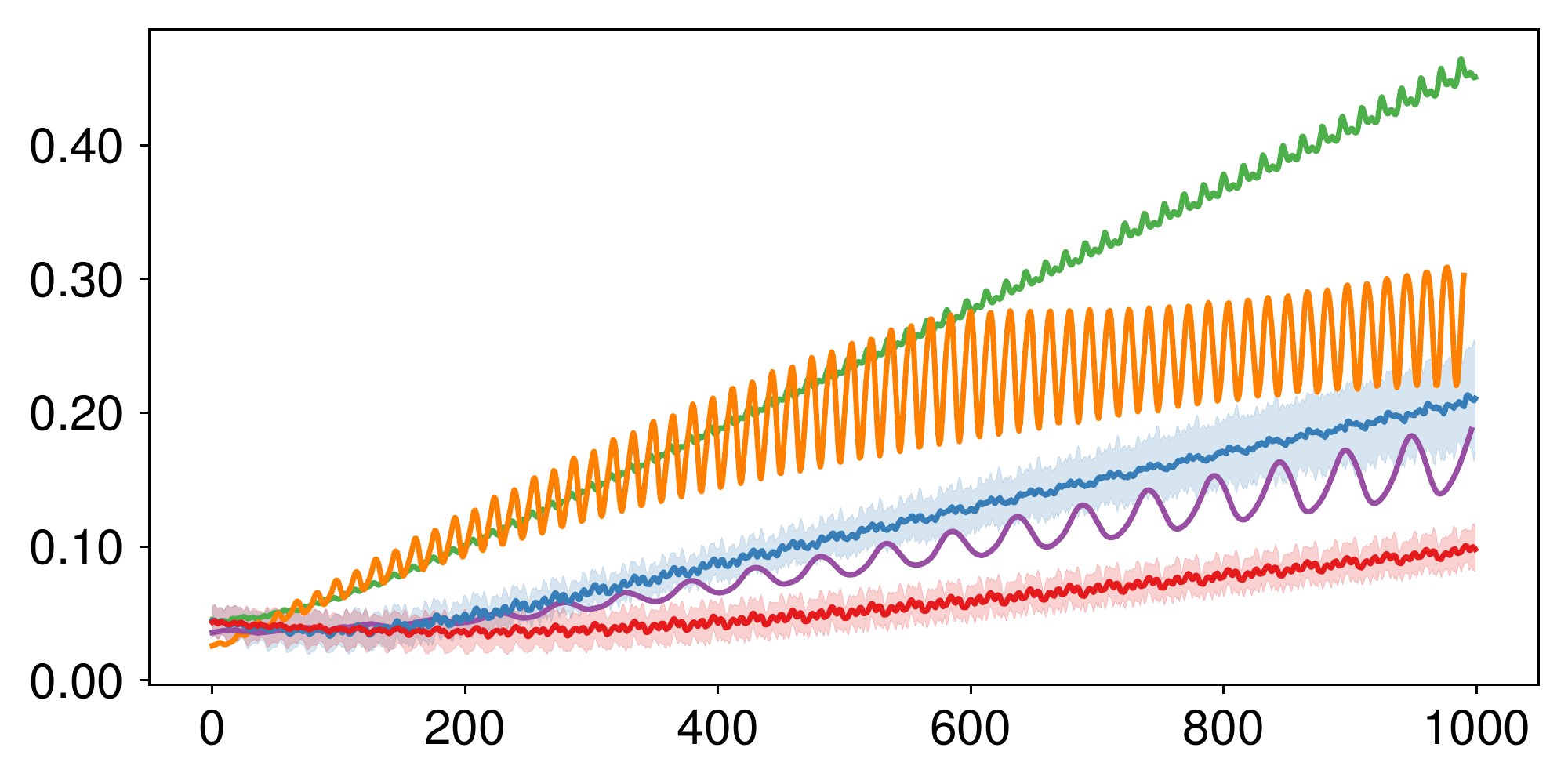} 
			\put(54,-4){\color{black}{\footnotesize $t$}}  		
		\end{overpic}\vspace{+0.32cm}			
		
		\caption{Noisy observations}\label{fig:pendulum_1_noise_c}
	\end{subfigure}\vspace{-0.2cm}
	
	\caption{Prediction errors, over a time horizon of $1000$ steps, for clean and noisy observations from a pendulum with initial conditions $\theta_0=0.8$ (top row) and $\theta_0=2.4$ (bottom row). Our model outperforms the DAE and the Hamiltonian NN in all settings.}
	\label{fig:pendulum_1_noise}
\end{figure*}

\subsection{Nonlinear Pendulum with no Friction}
\label{subsec:pendulum}

The nonlinear (undamped) pendulum~\cite{hirsch1974differential} is a classic textbook example for dynamical systems, which is also used for benchmarking deep models~\citep[e.g., ][]{greydanus2019hamiltonian,bertalan2019learning,chen2020symplectic}. This problem can be modeled as a second order ODE by
\begin{equation}
\frac {d^{2}\theta }{dt^{2}}+{\frac {g}{\ell }}\sin \theta = 0 \ ,
\end{equation}
where the angular displacement from an equilibrium is denoted by $\theta \in [0,2\pi)$. We use $l$ and $g$ to denote the length and gravity, respectively, with $l=1$ and $g=9.8$ in practice. We consider the following initial conditions $\theta(0) = \theta_0$ and $\dot{\theta}(0)=0$. The motion of the pendulum is approximately harmonic for a small amplitude of the oscillation $\theta_0 \ll 1$. However, the problem becomes inherently nonlinear for large amplitudes of the oscillations. 

We experiment with oscillation angles $\theta_0=0.8$ and $\theta_0=2.4$ on the time interval $t=[0,170]$. The data are generated using a time step $\Delta t = 0.1$, yielding $T = 1700$ equally spaced points ${x}_1, ..., {x}_T \in \mathbb{R}^2$. In addition, we map the sequence $\{ {x}_t\}$ to a high-dimensional space via a random orthogonal transformation to obtain the training snapshots, i.e., $\mathbf{P} \in \mathbb{R}^{64 \times 2}$ such that $f_t = \mathbf{P} \, {x}_t$ for any $t$. Finally, we split the new sequence into a training set of $600$ points and leave the rest for the test set. We show in Fig.~\ref{fig:ho_data} examples of the clean and noisy trajectories for these data.

\textbf{Experimental results.} \
Fig.~\ref{fig:pendulum_1_noise} shows the pendulum results for initial conditions $\theta_0=0.8$ (top row) and $\theta_0=2.4$ (bottom row). We used a bottleneck $\kappa=6$ and $\alpha=0.5$ for the DAE and our models. (The parameter $\alpha$ controls the width of the network, i.e., the number of neurons used for the hidden layers.) The relative forecasting error is computed at each time step via ${\|f_t - \hat{f}_t \|_2}/{\|f_t\|_2}$, where $\hat{f}_t$ is the high-dimensional estimated prediction, see also Eq.~\eqref{eq:approx_fwd_pred}. We forecast over a time horizon of $1000$ steps, and we average the error over $30$ different initial observations $f_t$, where the shaded areas represent the $\pm 1$ standard deviations. Overall, our model yields the best results in all the cases we explored. The RNN obtains good measures in the clean case and for short prediction times, but its performance deteriorates in the noisy setup and when forecasting is required for long horizons. The DAE model~\cite{lusch2018deep} recovers the pendulum dynamics in the linear regime but struggles when the nonlinearity increases. 
We also outperform the Hamiltonian NN~\cite{greydanus2019hamiltonian} in all settings.

Tab.~\ref{tab:results_pendulum} lists the prediction error at the final time step ${\|f_T - \hat{f}_T \|_2}/{\|f_T\|_2}$ for both the DAE and our model. For each case we list the minimum, maximum and average prediction error evaluated for models that were trained with $18$ different seed values.

\begin{table}[!h]
	\centering\scalebox{0.9}{
		\begin{tabular}{lcccccccc} \toprule
			\multicolumn{1}{c}{model} & \multicolumn{1}{c}{$\theta$} & \multicolumn{1}{c}{noise} & \multicolumn{3}{c}{prediction error} & \multicolumn{1}{c}{\#parms} \\
			& & & min & max & avg. & & \\ \midrule
			DAE & 0.8 & - & 0.016    &  0.254   &  0.102 & 0.03M\\
			Ours & 0.8 & - &  \textbf{0.011}  &  0.080   & 0.034 & 0.03M \\ \midrule
			DAE & 0.8 & true & 0.062    &  0.563   &  0.232 & 0.03M\\
			Ours & 0.8 & true & \textbf{0.045}  &  0.362   & 0.131 & 0.03M\\ \midrule
			
			DAE & 2.4 & - & 0.042    &  0.211   &  0.112 & 0.03M\\
			Ours & 2.4 & - &  \textbf{0.027}  &  0.171   & 0.074 & 0.03M \\ \midrule
			DAE & 2.4 & true & 0.191    & 0.521    &  0.316 & 0.03M\\
			Ours & 2.4 & true & \textbf{0.063}  &  0.592  & 0.187 & 0.03M\\					
			\bottomrule 
	\end{tabular}}
	\caption{Summary of results for the nonlinear pendulum with no friction. Our consistent Koopman model has a significantly reduced variance, and it outperforms the DAE in all situations. Each model was trained with $18$ different seed values. }	\label{tab:results_pendulum}
\end{table}

\subsection{High-dimensional Fluid Flows}
\label{subsec:fluid_flows}
	
Next, we consider two challenging fluid flow examples. The first instance is a periodic flow past a cylinder that exhibits vortex shedding from boundary layers. This flow is commonly used in physically-based machine learning studies \cite{takeishi2017learning,morton2018deep}. The data are generated by numerically solving the Navier--Stokes equations given here in their vorticity form $\partial_t \omega = - \langle v, \nabla \omega \rangle + \frac{1}{Re} \Delta \omega$, where $\omega$ is the vorticity taken as the curl of the velocity, $\omega = \mathrm{curl}(v)$. We employ an immersed boundary projection solver~\cite{taira2007immersed} with $Re=100$. Our simulation yields $300$ snapshots of $192\times 199$ grid points, sampled at regular intervals in time, spanning five periods of vortex shedding. We split the data in half for training and testing. Our second example is an inviscid flow, i.e., $Re=\infty$, of a vortex pair travelling over a curved domain of a sphere given as a triangle mesh with $2562$ nodes. We facilitate the intrinsic solver~\cite{azencot2014functional} by producing $600$ snapshots of which we use $550$ for training.

\begin{table}[!b]
	\centering\scalebox{0.9}{
		\begin{tabular}{lcccccccc} \toprule
			\multicolumn{1}{c}{model} & \multicolumn{1}{c}{noise} & \multicolumn{3}{c}{prediction error} & \multicolumn{1}{c}{\#parms} \\
			& & min & max & avg. & & \\ \midrule
			DAE & - & 0.078 & 0.348 & 0.156 & 2.49M\\
			Ours & - & 0.064 & 0.196 & 0.112 & 2.49M \\ \midrule
			DAE & true & 0.125 & 0.982 & 0.343 & 2.49M\\
			Ours & true & 0.114 & 0.218 & 0.151 & 2.49M\\ 			
			\bottomrule 
	\end{tabular}}
	\caption{Summary of results for the flow past the cylinder. Each model was trained with $18$ different seed values. }
	\label{tab:results_flow}
\end{table}

\textbf{Experimental results.}  \ 
It can be shown that the cylinder dynamics evolve on a low-dimensional attractor \cite{noack2003hierarchy}, which can be viewed as a nonlinear oscillator with a state-dependent damping~\cite{loiseau2018constrained}. As the dynamics are within the linear regime, we expect that on clean data, both DAE and ours will obtain good prediction results. We compare both models using width $\alpha=2$ and bottleneck $\kappa = 10$, and we list in Tab.~\ref{tab:results_flow} the obtained predictions at $T=100$. Indeed, for the clean data the results are similar in nature, whereas for the noisy case, our model provides more robust predictions. In general, our model outperforms the DAE when the data are noisy and the prediction horizon is long, illustrating the regularizing effect of our loss terms. 

In Tab.~\ref{tab:results_sphere_ns}, we show a few estimated predictions for the flow over a sphere. At the top part of the table we compare our method to DAE. Again, the results are consistent with previous examples where our approach yields better error and robustness estimates. In addition, we show a negative result of our method at the bottom of the table where we add increasing levels of diffusion to the Euler equation, making it non-invertible. Indeed, higher diffusion leads to inferior predictions when compared to the non-diffusive test.

\begin{table}[h]
	\centering\scalebox{0.9}{
		\begin{tabular}{lcccccccc} \toprule
			\multicolumn{1}{c}{model} & \multicolumn{1}{c}{diffusion} & \multicolumn{3}{c}{prediction error} & \multicolumn{1}{c}{\#parms} \\
			& & min & max & avg. & & \\ \midrule
			DAE & 0 & 0.071 & 1.558 & 0.715 & 0.34M\\
			Ours & 0 & 0.062 & 1.291 & 0.532 & 0.34M \\ \midrule
			Ours & 0.0001 & 0.123 & 1.122 & 0.349 & 0.34M \\
			Ours & 0.001 & 0.44 & 1.386 & 0.725 & 0.34M \\	
			Ours & 0.01 & 0.572 & 1.47 & 0.995 & 0.34M \\														
			\bottomrule 
	\end{tabular}}
	\caption{Summary of results for the flow over a sphere. Each model was trained with $18$ different seed values. }	\label{tab:results_sphere_ns}
\end{table}

\begin{figure*}[!t]\vspace{+0.2cm}
	\centering
	\begin{overpic}[width=0.95\textwidth]{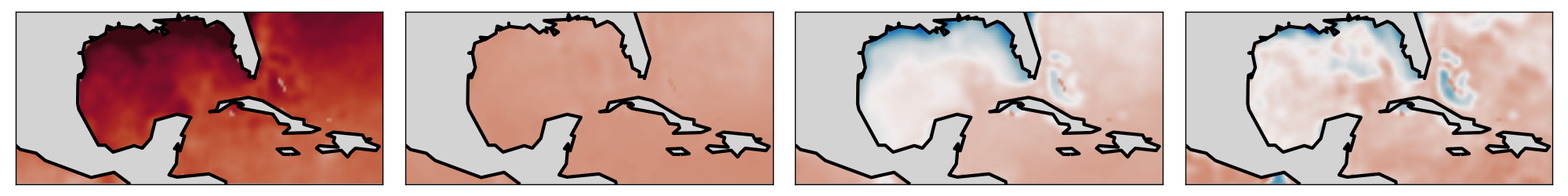} 
		\put(7,13){\footnotesize Initial condition}
		\put(35.5,13){\footnotesize DAE}
		\put(60,13){\footnotesize Ours}
		\put(82,13){\footnotesize Ground truth}
	\end{overpic}\vspace{-0.1cm}	
	
	\includegraphics[width=0.95\textwidth]{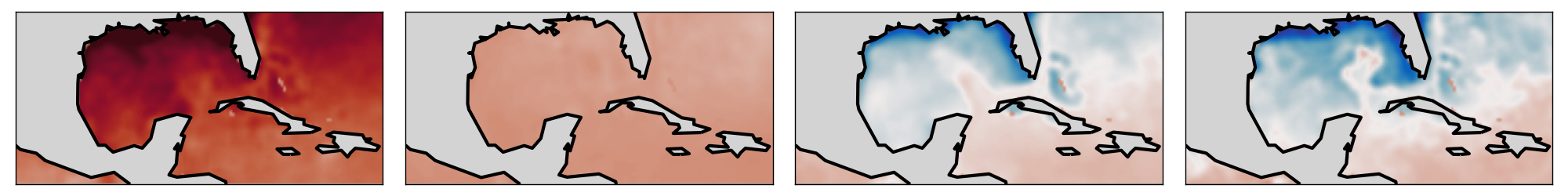}
	
	\caption{Starting from the same initial conditions, we use our model and DAE to forecast the sea surface temperature for the $120$th day (top row) and $175$th day (bottom row). The results can be visually compared to the ground truth data (right column). Our model generally attains predictions that are much closer to the ground truth compared to the results obtained by DAE.}
	\label{fig:sst_viz}
	\vspace{-2mm}
\end{figure*}

\subsection{Sea Surface Temperature Data}		

Our last dataset includes complex climate data representing the daily average sea surface temperature measurements around the Gulf of Mexico. This dataset is used in climate sciences to study the intricate dynamics between the oceans and the atmosphere. Climate prediction is generally a hard task involving challenges such as irregular heat radiation and flux as well as uncertainty in the wind behavior. Nevertheless, the input dynamics exhibit non-stationary periodic structures and are empirically low-dimensional, suggesting that Koopman-based methods can be employed. We extract a subset of the NOAA OI SST V2 High Resolution Dataset hereafter SST, and we refer to~\cite{reynolds2007daily} for additional details. In our experiments, we use data with a spatial resolution of $100\times 180$ spanning a time horizon of $1,305$ days, of which $1,095$ snapshots are used for training.

\textbf{Experimental results.} \
We demonstrate in Fig.~\ref{fig:sst_viz} a visual comparison of the estimated predictions as obtained from the DAE model and ours. Here we are using width $\alpha=6$ and bottleneck $\kappa = 10$. In both cases, forecasting for the $120$th day (top) and $175$th day (bottom), our results are closer to the ground truth by a large margin. A more detailed analysis is provided in Tab.~\ref{tab:results_sst} where we show the prediction errors averaged $\frac{1}{T}\sum_{t=1}^{T} |\hat{f}_t - f_t  | / | f_t|$ over a time horizon of $180$ days. (Here we use the average prediction error to better capture prediction errors that corresponds to seasonal effects.) 
The prediction results are significant, as predicting climate data is in general a challenging problem.

\begin{table}[!h]
	\centering\scalebox{0.9}{
		\begin{tabular}{lcccccccc} \toprule
			\multicolumn{1}{c}{model}  & \multicolumn{3}{c}{prediction error} & \multicolumn{1}{c}{\#parms} \\
			 & min & max & avg. & & \\ \midrule
			DAE  & 0.315 & 1.062 & 0.701 & 2.05M\\
			Ours  & 0.269 & 1.028 & 0.632 & 2.05M \\		
			\bottomrule 
	\end{tabular}}
	\caption{Summary of results for the sea surface temperature data. Each model was trained with $18$ different seed values. }
	\label{tab:results_sst}
\end{table}

\section{Ablation Study}

To support our empirical results, we have also conducted an ablation study to quantify the effect of our additional loss terms when weighted differently. To this end, we revisit the noisy pendulum flow with an initial condition $\theta_0=0.8$. Our results are summarized in Tab.~\ref{tab:Ablation} where we explore various values for $\lambda_\mathrm{bwd}$ and $\lambda_\mathrm{con}$ which balance the back prediction and consistency penalties, respectively.  Our model generally outperforms other baselines for all the parameters we checked, measured via the average error (4th column) and most distant prediction error (5th column).

\begin{table}[!h]
	\centering\scalebox{0.9}{
		\begin{tabular}{lcccccccc} \toprule
			Model	   &$\lambda_\mathrm{bwd}$     & $\lambda_\mathrm{con}$   & $\frac{1}{T}\sum_{t=1}^{T} \frac{||\hat{f}_t - f_t  ||_2}{ || f_t||_2}$        & $ \frac{||\hat{f}_T - f_T  ||_2}{ || f_T||_2}$   \\
			\midrule
			DAE		   & 0.0  &  0.0  			&  0.10 & 0.15 \\
			Ours 	&  $1\text{e-}1$  &  0.0   			&  0.06 &  0.11 \\
			\midrule
			Ours (*)	 &  $1\text{e-}1$  &  $1\text{e-}2$  &  0.06  	& 0.08 \\			
			Ours	 &  $1\text{e-}1$  &  $1\text{e-}1$ 	&  0.07  	&  0.12 \\		
			\midrule
			Ours	 &  $2\text{e-}1$  &  $1\text{e-}2$ 	&  \textbf{0.05}  	&  \textbf{0.06} \\
			Ours	 &  $2\text{e-}1$  &  $1\text{e-}1$ 	&  0.06  	&  0.08 \\
			\bottomrule 
	\end{tabular}}
	\caption{Ablation study for the pendulum with fixed seed 1. The star indicates the setting that is used during our experiments above. }
	\label{tab:Ablation}
\end{table}

\section{Discussion}

In this paper, we have proposed a novel physically constrained learning model for processing high-dimensional time series data. Our method is based on Koopman theory as we approximate dynamical systems via linear evolution matrices. Key to our approach is that we consider the backward dynamics during prediction, and we promote the consistency of the forward and backward systems. These modifications may be viewed as relaxing strict reversibility and stability constraints, while still regularizing the parameter space. We evaluate our method on several challenging datasets and compare with a state-of-the-art Koopman-based network as well as other baselines. Our approach notably outperforms the other models on noisy data and for long time predictions. We believe that our work can be extended in many ways, and in the future, we plan on considering our setup within a recurrent neural network design.

\section*{Acknowledgements}
We are grateful to the generous support from Amazon AWS.
OA would like to acknowledge the European Union's Horizon 2020 research and innovation programme under the Marie Sk{\l}odowska-Curie grant agreement No. 793800. NBE would like to acknowledge CLTC for providing support of this work. MWM would like to acknowledge ARO, DARPA, NSF and ONR for providing partial support of this work. We would like to thank J. Nathan Kutz, Steven L. Brunton, Lionel Mathelin and Alejandro Queiruga for valuable discussions about dynamical systems and Koopman theory. Further, we would like to acknowledge the NOAA for providing the SST data (\href{https://www.esrl.noaa.gov/psd/}{https://www.esrl.noaa.gov/psd/}).

\appendix

\section{Network architecture}

In our evaluation, we employ an autoencoding architecture where the encoder and decoder are shallow and contain only three layers each. Using a simple design allows us to focus our comparison on the differences between the DAE model~\cite{lusch2018deep} and ours. Specifically, we list the network structure in Tab.~\ref{tab:arch_model1} including the specific sizes we used as well as the different activation functions. We recall that $m$ represents the spatial dimension of the input signals, whereas $\kappa$ is the bottleneck of our approximated Koopman operators. Thus, $p=32\cdot\alpha$ is the main parameter with which we control the width and expressiveness of the autoencoder. We facilitate fully connected layers as some of our datasets are represented on unstructured grids. Finally, we note that the only difference between our net architecture and the DAE model is the additional backward linear layer.

\begin{table}[!ht]
	\centering\scalebox{0.8}{
		\begin{tabular}{lcccccccc} \toprule
			Type & Layer & Weight size & Bias size & Activation \\
			\midrule
			Encoder	& FC & $m \times p$ & $p$ & $\mathrm{tanh}$ \\
			Encoder	& FC & $p \times p$ &  $p$ & $\mathrm{tanh}$  \\
			Encoder	& FC & $p \times \kappa$ & $\kappa$ & linear \\
			\midrule
			Forward	& FC & $\kappa \times \kappa$ & $0$ & linear \\
			\textbf{Backward} & FC & $\kappa \times \kappa$ & $0$ & linear \\
			\midrule		
			Decoder	& FC & $\kappa \times p$ &  $p$ & $\mathrm{tanh}$ \\
			Decoder	& FC & $p \times p$ & $p$ & $\mathrm{tanh}$ \\
			Decoder	& FC & $p \times m$ & $m$ & linear \\ 
			\bottomrule 
	\end{tabular}}
	\caption{Our network architecture, where $p=16\cdot\alpha$ with $\alpha$ controlling the width per encoder and decoder layer.}
	\label{tab:arch_model1}
\end{table}

\section{Computational requirements}

The models used in this work are relatively shallow. The amount of parameters per model can be computed as follows $2(m + 32 + \kappa) \cdot 32 \alpha + (4 \cdot 32 \alpha + \kappa + m) + 2 \cdot \kappa^2$, corresponding to the number of weights, biases and Koopman operators, respectively. Notice that DAE is different than our model by having $\kappa^2$ less parameters.

\begin{figure}[!h]
	\centering
	\begin{overpic}[width=\linewidth]{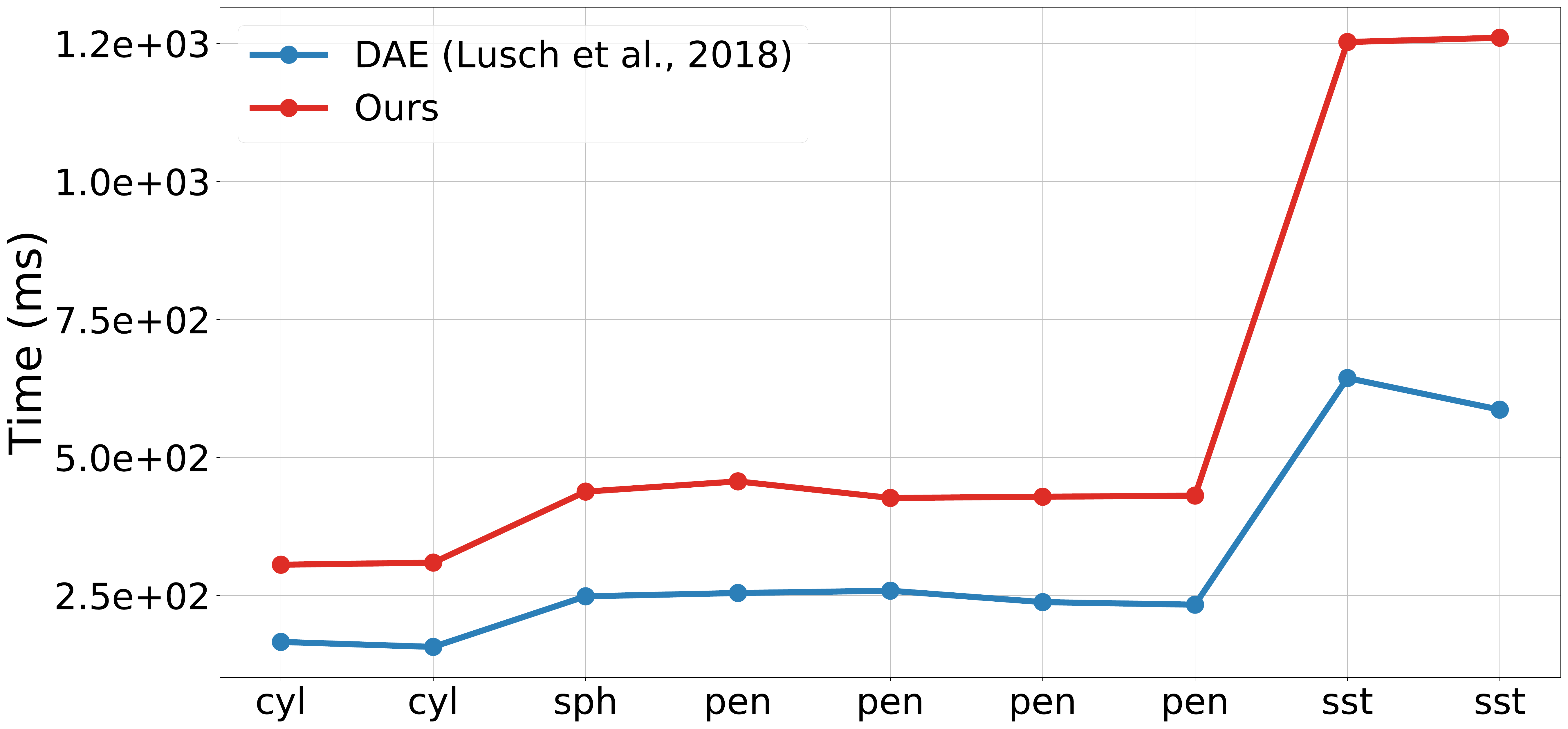}
	\end{overpic}
	\caption{We show above the average run time for an epoch in milliseconds for several of the test cases in this work. In general, our model is almost two times slower than DAE during training.}
	\label{fig:runtimes}
\end{figure}

We recorded the average training time per epoch, and we show the results for DAE and our models in Fig.~\ref{fig:runtimes} for many of our test cases. Specifically, the figure shows from left to right the run times for cylinder flow, noisy cylinder flow, sphere flow, linear pendulum, noisy linear pendulum, nonlinear pendulum, noisy nonlinear pendulum, SST and noisy SST. The behavior of our model is consistent in comparison to DAE for the different tests. On average, if DAE takes $x$ milliseconds per epoch, than our model needs $\approx 1.8x$ time. 

This difference in time is due to the additional penalty terms, and it can be asymptotically bounded by $$ \mathcal{O}(\mathcal{E}_\mathrm{bwd}) + \mathcal{O}(\mathcal{E}_\mathrm{con}) = \mathcal{O}(\lambda_s n m) + \mathcal{O}(\kappa^4) \ .$$ We note that the asymptotics for the forward prediction are equal to the backward component, i.e.,  $\mathcal{O}(\mathcal{E}_{\mathrm{fwd}})= \mathcal{O}(\lambda_s n m)$. The consistency term $\mathcal{E}_\mathrm{con}$ is composed of a sum of sequence of cubes (matrix products) which can be bounded by $\kappa^4$, assuming matrix multiplication is $\mathcal{O}(\kappa^3)$ and thus it is a non tight bound. Moreover, while the quartic bound is extremely high, we note that a cheaper version of the constraint can be used in practice, i.e.,  $||CD - I||_F^2$. Also, since our models are loaded to the GPU where matrix multiplication computations are usually done in parallel, the practical bound may be much lower. Finally, the inference time is insignificant ($\approx 1$ ms) and it is the same for DAE and ours and thus we do not provide an elaborated comparison. 

\section{Backward prediction of dynamical systems}

\begin{figure}[!h]
	\centering
	\begin{overpic}[width=\linewidth]{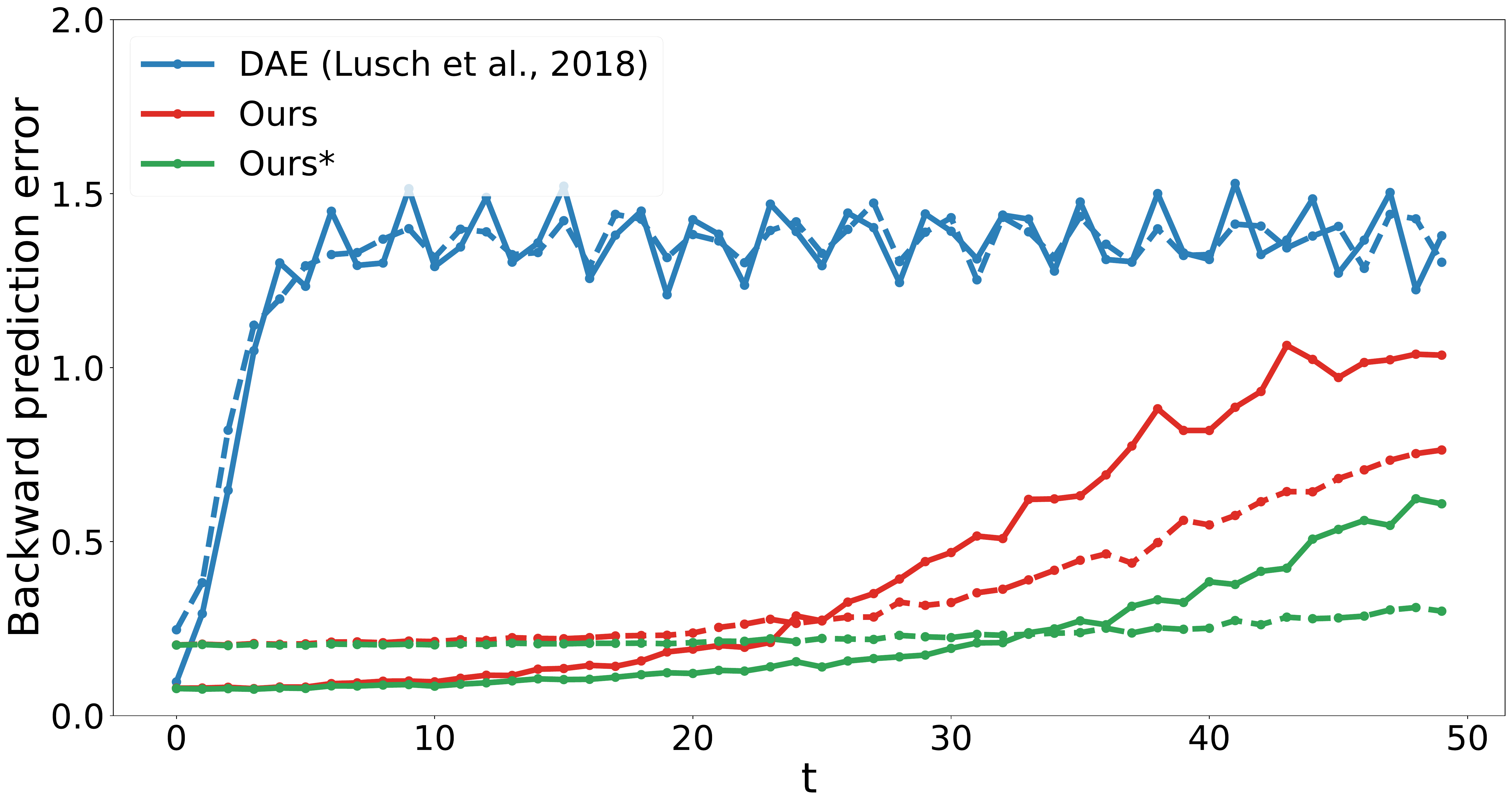}
	\end{overpic}
	\caption{The cylinder flow data is used for backward prediction with our model (red) and DAE (blue). Our results hint that DAE overfits in the forward direction, whereas our network generalizes well when the time is reversed.}
	\label{fig:back_predict}
\end{figure}

One of the key features of our model is that it allows for the direct backward prediction of dynamics. Namely, given an observation $f_t$, our network yields the forward prediction via $\hat{f}_{t+1} = \chi_d \circ C \circ \chi_e(f_t)$, as well as the backward estimate using $\check{f}_{t-1} = \chi_d \circ D \circ \chi_e(f_t)$. Time reversibility may be important in various contexts~\cite{greydanus2019hamiltonian}. For instance, given two different poses of a person, we can consider the trajectory from the first pose to the second or the other way around. Typically, neural networks require that we re-train the model in the reverse direction to be able to predict backwards. In contrast, Koopman-based methods can be used for this task as the Koopman matrix is linear and thus back forecasting can be obtained simply via $\bar{f}_{t-1} = \chi_d \circ C^{-1} \circ \chi_e(f_t)$. We show in Fig.~\ref{fig:back_predict} the backward prediction error computed with $\bar{f}_{t-1}$ for the cylinder flow data using our model and the DAE (blue and red curves). In addition, as our model computes the matrix $D$, we also show the errors obtained for $\check{f}_{t-1}$. The solid lines correspond to the clean version of the data, whereas the dashed lines are related to its noisy version. Overall, our model clearly outperforms DAE by an order of magnitude difference, indicating the overfitting in DAE.

\bibliographystyle{icml2020}
\bibliography{bib}

\end{document}